\documentclass{nature}

\usepackage[utf8]{inputenc}

\usepackage{graphicx}
\usepackage{color}
\usepackage[dvipsnames]{xcolor}

\usepackage{gensymb}
\usepackage{amssymb}
\usepackage{mathtools}

\usepackage{lineno}

\newcommand{\pb}{$^{210}$Pb\,}
\newcommand{\po}{$^{210}$Po\,}
\newcommand{\bi}{$^{210}$Bi\,}
\newcommand{\cpd}{cpd per 100\,t\,}
\newcommand{\cno}{CNO\,}

\newcommand{\PoMin}{(11.5 \pm 1.0)} 
\newcommand{\FinalSys}{ -0.5/+0.6  \,}
\newcommand{\FinalRate}{7.2\,}
\newcommand{\FinalSigmaL}{1.7\,}
\newcommand{\FinalSigmaR}{3.0\,}
\newcommand{\FinalSigma}{5.0}

\newcommand{\FinalFlux}{7.0\,}
\newcommand{\FinalFluxL}{2.0\,}
\newcommand{\FinalFluxR}{3.0\,}

\newcommand{\LondonCollege}{1}
\newcommand{\APC}{14}
\newcommand{\Dubna}{13}
\newcommand{\Genova}{9}
\newcommand{\Krakow}{18}
\newcommand{\Kiev}{19}
\newcommand{\Kurchatov}{3}
\newcommand{\Kurchatovb}{21}
\newcommand{\LNGS}{7}
\newcommand{\Milano}{5}
\newcommand{\Perugia}{23}
\newcommand{\Peters}{12}
\newcommand{\Princeton}{8}
\newcommand{\PrincetonChemEng}{6}
\newcommand{\UMass}{24}
\newcommand{\Virginia}{10}
\newcommand{\Munchen}{2}
\newcommand{\Lomonosov}{11}
\newcommand{\GSSI}{15}
\newcommand{\Dresda}{22}
\newcommand{\Mainz}{17}
\newcommand{\Juelich}{4}
\newcommand{\RWTH}{16}
\newcommand{\Aquila}{25}
\newcommand{\UniLondon}{20}

\newcommand{\spokes}{*}
\newcommand{\California}{26}
\newcommand{\Madrid}{27}
\newcommand{\LNGSG}{28}
\newcommand{\Napoli}{29}
\newcommand{\Hungary}{30}

\newcommand{\LondonCollegeText}{Department of Physics and Astronomy, University College London, Gower Street, London WC1E 6BT, UK}
 
\newcommand{\APCText}{Universit\'e de Paris, CNRS, Astroparticule et Cosmologie, F-75013 Paris, France}
\newcommand{\DubnaText}{Joint Institute for Nuclear Research, 141980 Dubna, Russia}
\newcommand{\GenovaText}{Dipartimento di Fisica, Universit\`a degli Studi and INFN, 16146 Genova, Italy}

\newcommand{\KrakowText}{M.~Smoluchowski Institute of Physics, Jagiellonian University, 30348 Krakow, Poland}
\newcommand{\KievText}{Institute for Nuclear Research of NAS Ukraine, 03028 Kyiv, Ukraine}
\newcommand{\KurchatovText}{National Research Centre Kurchatov Institute, 123182 Moscow, Russia}
\newcommand{\KurchatovbText}{National Research Nuclear University MEPhI (Moscow Engineering Physics Institute), 115409 Moscow, Russia}
\newcommand{\LNGSText}{INFN Laboratori Nazionali del Gran Sasso, 67010 Assergi (AQ), Italy}
\newcommand{\MilanoText}{Dipartimento di Fisica, Universit\`a degli Studi and INFN, 20133 Milano, Italy}
\newcommand{\PerugiaText}{Dipartimento di Chimica, Biologia e Biotecnologie, Universit\`a degli Studi e INFN, 06123 Perugia, Italy}
\newcommand{\PetersText}{St. Petersburg Nuclear Physics Institute NRC Kurchatov Institute, 188350 Gatchina, Russia}
\newcommand{\PrincetonText}{Physics Department, Princeton University, Princeton, NJ 08544, USA}
\newcommand{\PrincetonChemEngText}{Chemical Engineering Department, Princeton University, Princeton, NJ 08544, USA}
\newcommand{\UMassText}{Amherst Center for Fundamental Interactions and Physics Department, University of Massachusetts, Amherst, MA 01003, USA}
\newcommand{\VirginiaText}{Physics Department, Virginia Polytechnic Institute and State University, Blacksburg, VA 24061, USA}
\newcommand{\MunchenText}{ Physik-Department E15, Technische Universit\"at  M\"unchen, 85748 Garching, Germany}
\newcommand{\LomonosovText}{Lomonosov Moscow State University Skobeltsyn Institute of Nuclear Physics, 119234 Moscow, Russia}
\newcommand{\GSSIText}{Gran Sasso Science Institute, 67100 L'Aquila, Italy}
\newcommand{\DresdaText}{Department of Physics, Technische Universit\"at Dresden, 01062 Dresden, Germany}
\newcommand{\MainzText}{Institute of Physics and Excellence Cluster PRISMA$^+$, Johannes Gutenberg-Universit\"at Mainz, 55099 Mainz, Germany}

\newcommand{\JuelichText}{Institut f\"ur Kernphysik, Forschungszentrum J\"ulich, 52425 J\"ulich, Germany}
\newcommand{\RWTHText}{III. Physikalisches Institut B, RWTH Aachen University, 52062 Aachen, Germany}
\newcommand{\AquilaText}{Dipartimento di Scienze Fisiche e Chimiche, Universit\`a dell'Aquila, 67100 L'Aquila, Italy}

\newcommand{\NapoliText}{Present address: Dipartimento di Fisica, Universit\`a degli Studi Federico II e INFN, 80126 Napoli, Italy}

\newcommand{\LNGSGText}{Present address: INFN Laboratori Nazionali del Gran Sasso, 67010 Assergi (AQ), Italy}
\newcommand{\CaliforniaText}{Present address: University of California, Berkeley, Department of Physics, CA 94720, Berkeley, USA}
\newcommand{\UniLondonText}{Department of Physics, Royal Holloway, University of London, Department of Physics, School of Engineering, Physical and Mathematical Sciences, Egham, Surrey, TW20 OEX}
\newcommand{\MadridText}{Present address: Departamento de F\'{i}sica Te\'{o}rica, Universidad Aut\'{o}noma de Madrid, Campus Universitario de Cantoblanco, 28049 Madrid, Spain}
\newcommand{\spokesText}{Corresponding address: spokesperson-borex@lngs.infn.it}
\newcommand{\HungaryText}{Also  at Institute of Nuclear Research (Atomki), H-4001, Debrecen, POB.51.,  Hungary}

\title{Experimental evidence of neutrinos produced in the CNO fusion cycle in the Sun \\
\small The Borexino Collaboration\footnote{A list of participants and their affiliations appears at the end of the paper}}

\begin{document}

\maketitle

\begin{abstract}
For most of their existence stars are fueled by the fusion of hydrogen into helium proceeding via two theoretically well understood processes, namely the $pp$ chain and the CNO cycle~\cite{bib:SSM1,Vinyoles:2016djt}. 
Neutrinos emitted along such fusion processes in the solar core are the only direct probe of the deep interior of the star. 
A complete spectroscopy of neutrinos from the {\it pp} chain, producing about 99\% of the solar energy, has already been performed \cite{bib:Nature-2018}.  Here, we report the direct observation, with a high statistical significance, of neutrinos produced in the CNO cycle in the Sun. This is the first experimental evidence of this process obtained with the unprecedentedly radio-pure large-volume liquid-scintillator Borexino detector located at the underground Laboratori Nazionali del Gran Sasso in Italy. The main difficulty of this experimental effort is to identify the excess of the few counts per day per 100 tonnes of target due to CNO neutrino interactions above the backgrounds. 
A novel method to constrain the rate of \bi contaminating the scintillator relies on the thermal stabilisation of the detector achieved over the past 5 years. In the CNO cycle, the hydrogen fusion is catalyzed by the carbon (C) - nitrogen (N) – oxygen (O) and thus its rate, as well as the flux of emitted CNO neutrinos, directly depends on the abundance of these elements in solar core. Therefore, this result paves the way to a direct measurement of the solar metallicity by CNO neutrinos. While this result quantifies the relative contribution of the CNO fusion in the Sun to be of the order of 1\%, this process is dominant in the energy production of massive stars. The occurrence of the primary mechanism for the stellar conversion of hydrogen into helium in the Universe has been proven.
\end{abstract}


The nuclear fusion mechanisms active in stars, the {\it pp} chain and the CNO cycle, are associated with the production of energy and the emission of a rich spectrum of electron-flavour neutrinos\,\cite{bib:SSM1,Vinyoles:2016djt}, shown in Fig.\,1, lower plot. 
%
The relative importance of these mechanisms depends mostly on  stellar mass and on the abundance of elements heavier than helium in the core (``metallicity"). For stars similar to the Sun, but heavier than about 1.3\,$M_\odot$\,\cite{salaris}, the energy production rate is dominated by the CNO cycle,
while the {\it pp} chain prevails in lighter, cooler stars. 
The CNO cycle is believed to be the primary mechanism for the stellar conversion of hydrogen into helium in the Universe and is thought to contribute to the energy production in the Sun at the level of 1\%, with a large uncertainty related to poorly known metallicity. Metallicity is relevant for two reasons: $i$)  ``metals" directly catalyse the CNO cycle, and $ii$)  they affect the plasma opacity, indirectly changing  the temperature of the core and modifying the evolution of the Sun and its density profile. We notice 
that in the Sun 
the CNO sub-cycle I (see Fig.\,1, upper plot) is dominant~\cite{Angulo1999}. 

 The CNO neutrino flux scales with the metal abundance in the solar core, itself a tracer of the initial chemical composition of the Sun at the time of its formation. The metal abundance in the core is thought to be decoupled from the surface by a radiative zone, where no mixing occurs. CNO neutrinos are the only probe of that initial condition.

The neutrinos produced by the solar {\it pp} chain have been extensively studied since the early 70’s leading the discovery of nuclear fusion reactions in the Sun and the matter-enhanced neutrino flavour conversion\,\cite{bib:Homestake, bib:Gallex, bib:SAGE, bib:Art-Nobel2015,bib:SK, bib:SNO, bib:KamLAND}. Recently, the Borexino experiment has published a comprehensive study of the neutrino from the {\it pp} chain\,\cite{bib:Nature-2018}.

We report here the first direct detection of the \cno solar neutrinos and prove that the catalysed hydrogen fusion envisaged by Bethe and Weizs\"acker in the 30’s indeed exists\,\cite{bib:Bethe1939,bib:Weizsacker}. This result quantifies the rate of the CNO cycle in the Sun and paves the way to the solution of the long standing “solar metallicity
problem”\,\cite{Vinyoles:2016djt} 
arising from the discrepancy on the metallicity predicted by solar models using updated (low) metal abundances from spectroscopy (SSM-LZ)\,\cite{bib:LowZHighZ} and that inferred from helioseismology, which favors a higher metal content (SSM-HZ).
%
Despite detailed studies, this puzzle remains an open problem in solar physics.

The observation of CNO neutrinos reported here
experimentally confirms the overall solar picture and shows that a direct measurement of the metallicity of the Sun's core is within reach of an improved, future measurement.


\section*{Borexino detector and data}

Borexino is a solar neutrino experiment located underground at the Laboratori Nazionali del Gran Sasso in Italy, where the cosmic muon flux is suppressed by a factor of $\sim$10$^6$. The detector active core consists of approximately 280\,tonnes of liquid scintillator contained in a spherical nylon vessel of 4.25\,m radius. Particles interacting in the scintillator emit light that is detected by 2212 photomultiplier tubes~\cite{bib:DetPaper} (PMTs).

Solar neutrinos are detected in Borexino via elastic scattering off electrons. The total number of detected photons and their arrival times are used to reconstruct the electron recoil energy and the interaction point in the detector, respectively. The energy and spatial resolution in Borexino has slowly deteriorated over time due to the steady loss of PMTs  
 (on average 1238 channels are active for this analysis) and they are currently $\sigma_E/E \approx$6\% and $\sigma_{x,y,z} \approx$11\,cm for 1\,MeV events at the center of the detector.

The time profile of the scintillation light provides a powerful way to distinguish among different particle types ($\alpha$, $\beta^-$, and $\beta^+$) via pulse-shape discrimination methods\,\cite{bib:be7Long,bib:Nusol} and is essential for the selection of $^{210}$Po $\alpha$ decays used to constrain $^{210}$Bi background, as discussed below.

In spite of the very high number of solar neutrinos reaching the Earth ($\approx$6$\times$\,10$^{10}$ $ \nu$\,cm$^{-2}$\,s$^{-1}$) their interaction rate is low, namely few tens of counts per day (cpd) in 100\,tonnes (t) of scintillator. Their detection is especially challenging, because the neutrino signals cannot be easily disentangled from radioactive backgrounds. Borexino’s success rests on its unprecedented radio-purity 
combined with the careful selection of materials\,\cite{bib:ScienceTechnology2002} and clean assembly protocols. 

This paper is based on data collected during Borexino Phase-III, from July 2016 to February 2020, corresponding to 1072 days of live time. The event sample is filtered by applying a set of selection criteria\,\cite{bib:Nusol} \, that reduce events from residual radioactive impurities, cosmic muons, cosmogenic isotopes, instrumental noise, and external gamma rays. The latter are significantly suppressed by selecting events occurring within an innermost volume of scintillator (fiducial volume, FV) defined by a cut on the reconstructed radius and vertical position ($r$\,$<$\,2.8\,m and -1.8\,m$<$\,$z$\,$<$\,2.2\,m). Data are analysed in the electron recoil energy interval from 320 to 2640\,keV.

The counting rate of events surviving the selection as a function of their visible energy is shown in Fig.\,2. 
The data distribution is understood to be the sum of solar neutrino components and of backgrounds due to the decays of residual radioactive contaminants of the scintillator ($^{85}$Kr, $^{210}$Bi, $^{210}$Po, $^{40}$K) and of cosmogenic $^{11}$C, and to $\gamma$-rays from the decays of $^{40}$K, $^{214}$Bi, and $^{208}$Tl in the materials external to the scintillator.
These backgrounds have been characterized in Phase-II\,\cite{bib:Nusol} and their rates range between a few and tens of \cpd, to be compared with the expected \cno\, signal of a few \cpd. The key backgrounds for this study are $^{11}$C and $^{210}$Bi. Together with solar $pep$ neutrinos (an alternative proton fusion first step of the $pp$ chain), they represent the main obstacle to the extraction of the \cno signal, as discussed in the next section. The expected background due to the elastic scattering of $^{40}$K geo-antineutrinos\,\cite{geonu2019} is negligible. The yellow vertical band in Fig.\,2\,highlights the region of largest CNO signal-to-background ratio.

\section*{CNO neutrino detection strategy and the \bi \,challenge}

\label{CNO_strategy}

Neutrinos from the CNO cycle produce a broad energy spectrum ranging between 0 and 1740\,keV (see Fig.~1, lower plot). Consequently, the recoil energy  of electrons has a rather featureless continuous distribution that extends up to 1517\,keV (see Fig.~2). In this work, the three CNO neutrino components (Fig.~1) were treated as a single contribution by fixing the ratio between them according to the SSM prediction\,\cite{bib:SSM1,Vinyoles:2016djt}. Several backgrounds contribute to the same energy interval with a rate comparable to or larger than the signal.
To disentangle all contributions, we fit the data with a procedure similar to that adopted in \,\cite{bib:Nature-2018,bib:Nusol, bib:xfd} and described in Methods.

The CNO analysis is affected by two additional complications: the similarity between the CNO-$\nu$ recoil electron and the $^{210}$Bi $\beta^-$ spectra and their strong correlations  with the $pep$-$\nu$ recoil electron spectrum. In addition, the data are contaminated by cosmogenic $^{11}$C in the high energy part of the CNO spectrum. The muon-neutron-positron three-fold-coincidence (TFC) tagging technique\,\cite{bib:Nusol} for \, $^{11}$C is essential to make the \cno detection possible.

The sensitivity to CNO neutrinos is low unless the \bi and $pep$-$\nu$ rates are sufficiently constrained in the fit\,\cite{bib:sensitivity-paper}.
The $pep$-$\nu$ rate is constrained to 1.4\% precision \,\cite{bib:sensitivity-paper}, using: solar luminosity, robust assumptions on the $pp$ to $pep$ neutrino rate ratio, existing solar neutrino data\,\cite{bib:Vissani2019, bib:bergstrom}, and the most recent oscillation parameters\,\cite{bib:NuPars}.
We underline that the luminosity of the Sun depends very weakly on the contribution of the \cno cycle, making the $pep$ constraint essentially independent of any reasonable assumption on the \cno rate.

The other main background for the CNO-$\nu$ measurement comes from the decays of $^{210}$Bi\,\cite{bib:sensitivity-paper}, a $\beta$ emitter with a short half-life (5.013\,days) and whose decay rate is supported by $^{210}$Pb through the sequence:
\begin{equation} \label{pb-decay-chain}
^{210}\mathrm{Pb} \xrightarrow[22.3\, \mathrm{years}]{\beta^-}{^{210}}\mathrm{Bi} \xrightarrow[5\,\mathrm {days}]{\beta^-}{^{210}}\mathrm{Po} \xrightarrow[138.4\,\mathrm{days}]{\alpha} {^{206}}\mathrm{Pb}\,.
\end{equation}
We note that the endpoint energy of the \pb \,$\beta$-decay is 63.5\,keV, well below the analysis threshold (320\,keV). Therefore, 
the determination of the $^{210}$Bi content 
must rely on measuring $^{210}$Po\,\cite{Villante2011}. The $\alpha$ particles from \po decay, selected event-by-event by means of pulse-shape discrimination, are ideal tracers of \bi, if secular equilibrium in sequence (\ref{pb-decay-chain}) is achieved. It is hence crucial to understand under what conditions such an equilibrium is established. 

Since 2007, the data have shown that out-of-equilibrium components of \po were present in the fiducial volume. A dedicated effort was implemented to study and ultimately to prevent these components from migrating into the fiducial volume by stabilising the detector temperature. This upgrade allowed us to reach a sufficient equilibrium in one central sub-volume of the detector, which made the result reported in this paper possible. We distinguish between a {\it Scintillator} (S) $^{210}$Po component ($^{210}$Po${^{\mathrm{S}}}$) sourced by $^{210}$Pb in the liquid 
and assumed to be stable in time and in equilibrium with $^{210}$Bi, and a {\it Vessel} (V) component ($^{210}$Po${^{\mathrm{V}}}$).
The source of $^{210}$Po${^{\mathrm{V}}}$ for this dataset is understood to be $^{210}$Pb deposited on the inner surfaces of the vessel.
The daughter $^{210}$Po may detach and move into the scintillator by diffusion or following slow convective currents. 
It is important to note that, as explained in detail below, there is no evidence of $^{210}$Pb itself leaching from those surfaces, since the rate of $^{210}$Bi observed in the scintillator has not significantly changed over several years.

The diffusion length of \po \,atoms in one half-life  
is significantly less than the separation between the vessel and the FV (approximately 1\,m). We can therefore conclude that diffusion is negligible for both \po \,and  \bi. However, Borexino data show that slow convective currents caused by temperature gradients and variations may indeed carry the \po \,into the FV. The same effect does not occur for the short-lived \bi that might also detach from the vessel, since it decays before reaching the FV.

Prior to 2016, Borexino was equipped with neither detailed temperature mapping, thermal insulation, or active temperature control. Convective currents were substantial, because of seasonal temperature variations and human activities affecting the temperature of the experimental hall. The large fluctuations of the $^{210}$Po \,activity in the FV induced by these currents are shown in Fig.~3, where the \po \,rate in different detector positions is plotted as a function of time. It is evident that before 2016 the \po \,counts in the FV were both high ($>100$\,cpd per 100\,t) and greatly unstable, on time scales shorter than the \po \,half-life, because of sizeable fluid movements, which prevented the separation of Po$^{{\mathrm{S}}}$ from Po$^{{\mathrm{V}}}$.

In order to suppress convection, a stable vertical thermal gradient needed to be established. The Borexino installation atop a cold floor in contact with the rock, acting as an infinite thermal sink, offers a unique opportunity to achieve such a gradient, once the detector is insulated against instabilities of the air temperature. Thermal insulation of the detector was completed in December 2015 and an active temperature control system\,\cite{BravoBerguo2018} was installed in January 2016 atop the detector (see Methods).
A residual seasonal modulation of the order of 0.3$^\circ$C/6 months is still visible in the detector and in the rock below it, but its effect is small for the purpose of this paper. 

This extensive stabilisation effort paid off: the \po \,rate initially decreased and reached its lowest value in a region that we named \emph{Low Polonium Field} (LPoF), above the equator around $z \simeq +80$\,cm. The existence of this volume, compatible in size and location with fluid dynamics simulation\,\cite{bib:PoSimulation}, is crucial in determining the \bi\ constraint.
We note that the result of this paper is stable against small variations of the shape and location of the LPoF.

\section*{\bi constraint}
\label{Bismuth}

The amount of \bi \,in the scintillator is determined from the minimum value of the \po \,rate in the LPoF through the relation
\begin{equation} \label{eq:pomin}
R(^{210}{\rm Po}_{\rm{min}})\,=\,R(^{210}{\rm Bi})\,+\,R(^{210}{\rm Po}^{\rm{V}}),
\end{equation} 
where, as discussed above, the \bi \,rate is equal
to $^{210}$Po$^{\rm{S}}$ according to 
secular equilibrium. Since $^{210}{\rm Po}^{\rm{V}}$ is always positive,  $^{210}{\rm Po}_{\rm{min}}$ yields an upper limit for the \bi \,rate. 

The \po \,content is not spatially uniform within the LPoF but exhibits a clear minimum with no sizable plateau around it. This yields a robust upper limit for the rate of $^{210}{\rm Bi}$, but does not guarantee that $^{210}{\rm Po}^{\rm{V}}$ is actually zero.
Only a spatially extended minimum of the \po\, rate would have yielded a measurement of the \bi \,rate.

The minimum \po \,rate was estimated from the \po \,distribution within the LPoF with 2D and 3D fits following two mutually compatible procedures (see Methods). 
The spatial position of the minimum is stable over the analysis period (it slowly moves by less than 20\,cm per month), showing that the detector is in a fluid dynamical quasi-steady condition and that the \po \,rate minimum is not a statistical fluctuation. Both procedures consistently yield $R(^{210}{\rm Po}_{\rm{min}}) = \PoMin$ cpd per 100\,t. The error includes the systematic uncertainty of the fit (see Methods). 

The \bi \,rate can then be extrapolated over the whole FV, provided that it
is uniform in the FV during the time period over which the estimation is performed. 
Lacking the possibility to individually tag \bi \,events, the analysis is performed by selecting $\beta$-like events at energies where the relative bismuth contribution is maximum.
We find the \bi \, angular and spatial distribution 
uniform within errors.
The systematic uncertainty associated to possible spatial non-uniformity of \bi \,is conservatively estimated at 0.78\,cpd per 100\,t. The observed \bi uniformity in Phase-III is expected from the substantial fluid mixing that has occurred prior to the thermal insulation, and agrees with 2D and 3D fluid dynamic simulations.
%

Because of the low velocity of convection currents, the uniformity of \bi\ provides a convincing evidence that \pb does not leach off the vessel.
As a cross check, 
the rate of $\beta$-like events shows the expected 3.3\% annual modulation of the solar neutrino rate (dominated by $^7$Be-$\nu$) due to the eccentricity of the Earth's orbit, proving that background $\beta$-like events are stable in time.
See Methods for details.

In summary, the \bi \,rate used as a constraint in the CNO-$\nu$ analysis is:
\begin{equation} \label{eq:penalty}
R(^{210}{\rm Bi}) \leq (11.5 \pm 1.3)\,\, {\rm cpd~ per~100~t},
\end{equation}
which includes the statistical and systematic uncertainties in the \po \,minimum determination and the systematic uncertainty related to the \bi \,uniformity hypothesis (added in quadrature).

\section*{Results and conclusions}
\label{Results}

We performed a multivariate analysis, simultaneously fitting the energy spectra in the window between 320\,keV and 2640\,keV and the radial distribution of the selected events, with details given in Methods. The following rates are treated as free parameters: CNO neutrinos, $^{85}$Kr, $^{11}$C, internal and external $^{40}$K, external $^{208}$Tl and $^{214}$Bi, and $^{7}$Be neutrinos.  The $pep$ neutrino rate is constrained to ($2.74 \pm 0.04$)\,cpd per 100\,t by multiplying the standard likelihood with a symmetric Gaussian term. 
The upper limit to the \bi \,rate obtained from eq.\,\ref{eq:penalty} is enforced asymmetrically by multiplying the likelihood with a half-Gaussian term, i.e., leaving the \bi \,rate unconstrained between 0 and 11.5\,\cpd.

The reference spectral and radial distributions (PDFs) of each signal and background species to be used in the multivariate fit are obtained with a complete \textsc{Geant4}-based Monte Carlo simulation\,\cite{bib:Nusol,bib:MCPaper}.
The results of the multivariate fit for data in which the $^{11}$C has been subtracted with the TFC technique are shown in Fig.~2. 
The $p$-value of the fit is high (0.3) demonstrating the good agreement between data and the underlying fit model. 
The corresponding negative log-likelihood for CNO-$\nu$, profiled over the other neutrino fluxes and background sources, is shown in Fig.~4 (dashed black line in the right panel). The best fit value is 7.2\,cpd per 100\,t with an asymmetric confidence interval of -1.7\,cpd per 100\,t and +2.9\,cpd per 100\,t (68\% C.L., statistical error only), obtained from the quantile of the likelihood profile.

We have studied possible sources of systematic error following an approach similar to the one used in\,\cite{bib:Nature-2018,bib:Nusol}. We have investigated the impact of varying fit parameters (fit range and binning) on the result by performing 2500 fits in different conditions and found it to be negligible with respect to the CNO statistical uncertainty.
We also considered the effect of different theoretical $^{210}$Bi shapes from\,\cite{bib:Daniel1962,bib:Carles1996,bib:210BiDerbin} and found that the CNO result is robust with respect to the selected one\,\cite{bib:Daniel1962}. Differences are included in the systematic error. We have performed a detailed study of the impact of possible deviations of the energy scale and resolution from the Monte Carlo model: non-linearity, non-uniformity, and variation in the absolute magnitude of the scintillator light yield have been investigated by simulating several million Monte Carlo pseudo-experiments with deformed shapes and fitting them with the regular non-deformed PDFs. The magnitude of the deformations was chosen to be within the range allowed by the available calibrations\,\cite{bib:bxcalib} and by two "standard candles" ($^{210}$Po, $^{11}$C) present in the data. The overall contribution to the total error of all these sources is \FinalSys\,cpd per 100\,t. Folding the systematic uncertainty over the log-likelihood profile we determine the final CNO interaction rate to be $\FinalRate^{+\FinalSigmaR}_{-\FinalSigmaL}$ \,cpd per 100\,t.
The rate can be converted to a flux of CNO-$\nu$ on Earth of \FinalFlux$^{+\FinalFluxR}_{-\FinalFluxL}\times10^8$\,cm$^{-2}$\,s$^{-1}$, assuming MSW conversion in matter\,\cite{deHolanda2004}, neutrino oscillation parameters from\,\cite{Capozzi2018} and Refs. therein, and a density of electrons in the scintillator of (3.307\,$\pm$\,0.015)\,$\times$\, 10$^{31}$\,e$^-$ per 100\,t.

Other sources of systematic error investigated in the previous precision measurement of the {\it pp} chain\,\cite{bib:Nature-2018}, such as, fiducial volume, scintillator density, and lifetime were found to be negligible with respect to the large CNO statistical uncertainty.

The log-likelihood profile including all the errors combined in quadrature is shown in Fig.~4 right (black solid line). The asymmetry of the profile is due to the applied
half-Gaussian constraint on the \bi, see eq.\,(\ref{eq:penalty}) and causes the profile to be relatively steep on the left side of the minimum. The shallow shape on the right side of the profile reflects the modest sensitivity to distinguish the spectral shapes of \bi \,and CNO recoil spectra.
 From the corresponding profile-likelihood we obtain a 5.1$\sigma$ significance of the CNO observation.
Additionally, a hypothesis test based on a profile likelihood test statistic\,\cite{bib:cowan} using 13.8\,million pseudo-data sets with the same exposure of Phase-III and systematic uncertainties included (see Methods) excludes the no-CNO signal scenario with a significance better than \FinalSigma$\sigma$ at 99.0\% C.L.

This observed CNO rate is compatible with both SSM-HZ and SSM-LZ predictions. Thus we cannot distinguish between the two different models (we quote the statistical compatibility for the reader: HZ is at 0.5$\sigma$ and LZ is at 1.3$\sigma$), see Fig.~4. When combined with other solar neutrino fluxes measured by Borexino the LZ hypothesis is disfavoured at 2.1$\sigma$.

We underline that the sensitivity to CNO neutrinos mainly comes from a small energy region between 780\,keV and 885\,keV (region of interest, ROI, see yellow band in Fig.~2), where the signal-to-background ratio is maximal\,\cite{bib:sensitivity-paper}. In this region, the count rate is dominated by events from CNO and $pep$ neutrinos, and by $^{210}$Bi decays. The remaining backgrounds contribute less than 20\% (see Fig.~4, left plot). A simple counting analysis confirms that the number of events in the ROI exceeds the sum from all known backgrounds, leaving room for CNO neutrinos, as depicted in Fig.~4, left. In this simplified approach (described in detail in Methods), we use the  $^{210}$Bi rate as determined in eq.\,\ref{eq:penalty} applying a symmetric Gaussian penalty and assuming an analytical description of the background model and the detector response. The statistical significance of the presence in the data of CNO neutrino events from this counting analysis, after accounting for statistical and systematic errors is lower  ($\simeq$\,$3.5\sigma$) than that obtained with the main analysis, given the simplified nature of this approach. 
In conclusion, we exclude the absence of a CNO solar neutrino signal with a significance of \FinalSigma$\sigma$. 
This is the first direct detection of CNO solar neutrinos.



\begin{addendum}
 \item We acknowledge the generous hospitality and support of the Laboratori Nazionali del Gran Sasso (Italy). The Borexino program is made possible by funding from Istituto Nazionale di Fisica Nucleare (INFN) (Italy), National Science Foundation (NSF) (USA), Deutsche Forschungsgemeinschaft (DFG) and Helmholtz-Gemeinschaft (HGF) (Germany), Russian Foundation for Basic Research (RFBR) (Grants No. 16-29-13014ofi-m, No. 17-02-00305A, and No. 19-02-00097A) and Russian Science Foundation (RSF) (Grant No. 17-12-01009) (Russia), and Narodowe Centrum Nauki (NCN) (Grant No. UMO 2017/26/M/ST2/00915) (Poland).
We gratefully acknowledge the computing services of Bologna INFN-CNAF data centre and U-Lite Computing Center and Network Service at LNGS (Italy), and the computing time granted through JARA on the supercomputer JURECA\,\cite{jureca} at Forschungszentrum J\"ulich (Germany). This research was supported in part by PLGrid Infrastructure (Poland).
\item[Authors Contributions] The Borexino detector was designed, constructed, and commissioned by the Borexino Collaboration over the span of more than 30 years. The Borexino Collaboration sets the science goals. Scintillator purification and handling, material radiopurity assay, source calibration campaigns, photomultiplier tube and electronics operations, signal processing and data acquisition, Monte Carlo simulations of the detector, and data analyses were performed by Borexino members, who also discussed and approved the scientific results. This manuscript was prepared by a subgroup of authors appointed by the Collaboration and subjected to an internal collaboration-wide review process. All authors reviewed and approved the final version of the manuscript.
 \item[Competing Interests] The authors declare that they have no
competing financial interests.
 \item[Correspondence] Correspondence and requests for materials
should be addressed to the Borexino Collaboration through the spokesperson's email: spokesperson-borex@lngs.infn.it. \item[Data Availability] The datasets generated during the current study are freely available in the repository https://bxopen.lngs.infn.it/. Additional information is available from the Borexino Collaboration spokesperson (spokesperson-borex@lngs.infn.it) upon reasonable request.
\end{addendum}


The Borexino Collaboration\textsuperscript{\spokes}

M.~Agostini\textsuperscript{\LondonCollege, \Munchen},
K.~Altenm\"{u}ller\textsuperscript{\Munchen},
S.~Appel\textsuperscript{\Munchen},
V.~Atroshchenko\textsuperscript{\Kurchatov},
Z.~Bagdasarian\textsuperscript{\Juelich,\California},
D.~Basilico\textsuperscript{\Milano},
G.~Bellini\textsuperscript{\Milano},
J.~Benziger\textsuperscript{\PrincetonChemEng},
R.~Biondi\textsuperscript{\LNGS},
D.~Bravo\textsuperscript{\Milano,\Madrid},
B.~Caccianiga\textsuperscript{\Milano},
F.~Calaprice\textsuperscript{\Princeton},
A.~Caminata\textsuperscript{\Genova},
P.~Cavalcante\textsuperscript{\Virginia,\LNGSG},
A.~Chepurnov\textsuperscript{\Lomonosov},
D.~D'Angelo\textsuperscript{\Milano},
S.~Davini\textsuperscript{\Genova},
A.~Derbin\textsuperscript{\Peters},
A.~Di Giacinto\textsuperscript{\LNGS},
V.~Di Marcello\textsuperscript{\LNGS},
X.F.~Ding\textsuperscript{\Princeton},
A.~Di Ludovico\textsuperscript{\Princeton},
L.~Di Noto\textsuperscript{\Genova},
I.~Drachnev\textsuperscript{\Peters},
A.~Formozov\textsuperscript{\Dubna,\Milano},
D.~Franco\textsuperscript{\APC},
C.~Galbiati\textsuperscript{\Princeton,\GSSI},
C.~Ghiano\textsuperscript{\LNGS},
M.~Giammarchi\textsuperscript{\Milano},
A.~Goretti\textsuperscript{\Princeton,\LNGSG},
A.S. G\"{o}ttel\textsuperscript{\Juelich,\RWTH},
M.~Gromov\textsuperscript{\Lomonosov,\Dubna},
D.~Guffanti\textsuperscript{\Mainz},
Aldo~Ianni\textsuperscript{\LNGS},
Andrea~Ianni\textsuperscript{\Princeton},
A.~Jany\textsuperscript{\Krakow},
D.~Jeschke\textsuperscript{\Munchen},
V.~Kobychev\textsuperscript{\Kiev},
G.~Korga\textsuperscript{\UniLondon,\Hungary},
S.~Kumaran\textsuperscript{\Juelich,\RWTH},
M.~Laubenstein\textsuperscript{\LNGS},
E.~Litvinovich\textsuperscript{\Kurchatov,\Kurchatovb},
P.~Lombardi\textsuperscript{\Milano},
I.~Lomskaya\textsuperscript{\Peters},
L.~Ludhova\textsuperscript{\Juelich,\RWTH},
G.~Lukyanchenko\textsuperscript{\Kurchatov},
L.~Lukyanchenko\textsuperscript{\Kurchatov},
I.~Machulin\textsuperscript{\Kurchatov,\Kurchatovb},
J.~Martyn\textsuperscript{\Mainz},
E.~Meroni\textsuperscript{\Milano},
M.~Meyer\textsuperscript{\Dresda},
L.~Miramonti\textsuperscript{\Milano},
M.~Misiaszek\textsuperscript{\Krakow},
V.~Muratova\textsuperscript{\Peters},
B.~Neumair\textsuperscript{\Munchen},
M.~Nieslony\textsuperscript{\Mainz},
R.~Nugmanov\textsuperscript{\Kurchatov,\Kurchatovb}
L.~Oberauer\textsuperscript{\Munchen},
V.~Orekhov\textsuperscript{\Mainz},
F.~Ortica\textsuperscript{\Perugia},
M.~Pallavicini\textsuperscript{\Genova},
L.~Papp\textsuperscript{\Munchen},
L.~Pelicci\textsuperscript{\Milano},
\"O.~Penek\textsuperscript{\Juelich,\RWTH},
L.~Pietrofaccia\textsuperscript{\Princeton},
N.~Pilipenko\textsuperscript{\Peters},
A.~Pocar\textsuperscript{\UMass},
G.~Raikov\textsuperscript{\Kurchatov},
M.T.~Ranalli\textsuperscript{\LNGS},
G.~Ranucci\textsuperscript{\Milano},
A.~Razeto\textsuperscript{\LNGS},
A.~Re\textsuperscript{\Milano},
M.~Redchuk\textsuperscript{\Juelich,\RWTH},
A.~Romani\textsuperscript{\Perugia},
N.~Rossi\textsuperscript{\LNGS},
S.~Sch\"onert\textsuperscript{\Munchen},
D.~Semenov\textsuperscript{\Peters},
G. Settanta\textsuperscript{\Juelich},
M.~Skorokhvatov\textsuperscript{\Kurchatov,\Kurchatovb},
A.~Singhal\textsuperscript{\Juelich,\RWTH},
O.~Smirnov\textsuperscript{\Dubna},
A.~Sotnikov\textsuperscript{\Dubna},
Y.~Suvorov\textsuperscript{\LNGS,\Kurchatov,\Napoli},
R.~Tartaglia\textsuperscript{\LNGS},
G.~Testera\textsuperscript{\Genova},
J.~Thurn\textsuperscript{\Dresda},
E.~Unzhakov\textsuperscript{\Peters},
F.L.~Villante\textsuperscript{\LNGS,\Aquila},
 A.~Vishneva\textsuperscript{\Dubna},
R.B.~Vogelaar\textsuperscript{\Virginia},
F.~von~Feilitzsch\textsuperscript{\Munchen},
M.~Wojcik\textsuperscript{\Krakow},
M.~Wurm\textsuperscript{\Mainz},
S.~Zavatarelli\textsuperscript{\Genova},
K.~Zuber\textsuperscript{\Dresda},
G.~Zuzel\textsuperscript{\Krakow}.

\noindent
\textsuperscript{*}\spokesText

\begin{affiliations}
\item \LondonCollegeText;
\item \MunchenText;
\item \KurchatovText;
\item \JuelichText;
\item \MilanoText;
\item \PrincetonChemEngText;
\item \LNGSText;
\item \PrincetonText;
\item \GenovaText;
\item \VirginiaText;
\item \LomonosovText;
\item \PetersText; 
\item \DubnaText; 
\item \APCText; 
\item \GSSIText; 
\item \RWTHText; 
\item \MainzText; 
\item \KrakowText; 
\item \KievText; 
\item \UniLondonText; 
\item \KurchatovbText; 
\item \DresdaText; 
\item \PerugiaText; 
\item \UMassText; 
\item \AquilaText. 
\item \CaliforniaText;
\item \MadridText; 
\item \LNGSGText; 
\item \NapoliText; 
\item \HungaryText.
\end{affiliations}


\begin{methods}

\section*{Experimental setup and neutrino detection technique}

The Borexino detector\,\cite{bib:DetPaper} was designed and built to achieve the utmost radio-purity at its core. It is made of an unsegmented Stainless Steel Sphere (SSS) mounted within a large Water Tank (WT). The SSS contains the organic liquid
and supports the photomultipliers (PMTs), while the water shields the SSS against external radiation and is the active medium of a Cherenkov muon tagger. A schematic drawing is shown in ED Fig.~1. 

Within this SSS, two thin (125\,$\mu$m) nylon vessels separate the volume in three shells of radii 4.25\,m, 5.50\,m, and 6.85\,m, the latter being the radius of the SSS itself. 

The inner nylon vessel (IV), concentric to the SSS, contains a solution of Pseudocumene (PC) as solvent and PPO (2,5-diphenyloxazole) as fluor dissolved at a concentration of about 1.5\,g/l. The second and the third shells are filled with a buffer liquid comprised of a solution of DMP (dimethylphthalate) in PC. The purpose of this double buffer is to shield the IV against $\gamma$ radiation emitted by contaminants present in the PMTs and the steel, while the outer nylon vessel prevents the diffusion of emanated Radon into the IV. The total amount of liquid within the SSS is approximately 1300\,tonnes, of which about 280\,tonnes are the active liquid scintillator. 

The IV scintillator density is slightly smaller than that of the buffer liquid, yielding an upward buoyant force. The IV is therefore anchored to the bottom of the SSS through thin high molecular weight polyethylene cords, thus minimising the amount of material close to the scintillator and keeping the IV in stable mechanical equilibrium.

The SSS is equipped with nominally 2212 8" PMTs that collect scintillation light emitted when a charged particle, either produced by neutrino interactions or by radioactivity, releases energy in the scintillator. Most of the PMTs (1800) are equipped with light concentrators (Winston cones) for an effective optical coverage of 30\%. Scintillation light is detected at  approximately 500\,photoelectrons per MeV of electron equivalent of deposited energy (normalized to 2000 PMTs). 
In organic liquid scintillators, the light yield per unit of deposited energy is affected by ionisation quenching\,\cite{birks64}. Alpha particles, characterised by higher ionisation rates along their path, experience more quenching compared to electrons and thus, produce less scintillation light.
The distribution of photon arrival times on PMTs allows the reconstruction of the location of the energy deposit by means of time-of-flight triangulation and the determination of the particle type by exploiting the pulse shape\,\cite{bib:Nusol}.

The very nature of the scintillation emission makes it impossible to distinguish the signal emitted by electrons scattered by neutrinos from that produced by electrons emitted in nuclear $\beta$ decays or Compton-scattered by $\gamma$ rays. Therefore, the radioactive background must be kept at or below the level of the expected signal rate, which for the total solar neutrino spectrum is of the order of a few events per tonne per day and, in the case of CNO neutrinos, two orders of magnitude smaller. Taking into account that typical materials (air, water, metals) are normally contaminated with radioactive impurities at the level of 10,000 or even 100,000 decays per tonne per second, this requirement is indeed a formidable challenge.

The scintillator procurement procedure was conceived to select an organic hydrocarbon with a very low $^{14}$C($\beta^-$, $Q$\,=\,156\,keV) content. Carbon-14 is cosmogenically activated in atmospheric carbon and an irreducible radioactive contaminant in organic hydrocarbons. The scintillator was delivered to the Gran Sasso laboratory in special tanks following procedures conceived to avoid contamination and to minimise the exposure to cosmic rays, which also produce other long living isotopes. Once underground, it was purified following various steps in plants specifically developed over more than 10 years for this purpose  and installed close to the detector. The purification during the 2007 initial scintillator fill was done mainly by distillation and counterflow sparging using low-argon-krypton nitrogen. 
A dedicated purification campaign in 2010\,-\,2011 processed the scintillator through several cycles of ultra-pure water extraction. These purification techniques are described in\,\cite{bib:PurifPlants},\,\cite{bib:FluidHandling}, and\,\cite{bib:Nusol}. 

This effort paid off: the extreme purity of the scintillator and the careful selection of the material surrounding it (nylon, plastic supports of the nylon vessels, steel, and PMT glass in particular), the use of carefully selected components (valves, pumps, fittings, etc.) together with special care during detector construction and installation yielded unprecedented low values of radioactive contaminants in the active scintillator. In addition, through the selection of a fiducial volume, the residual external gamma ray background (from IV nylon, SSS, and PMTs) is substantially reduced further. All Borexino results owe directly to this unprecedented radio-purity. 

The Water Tank is itself equipped with 208 PMTs to detect Cherenkov light emitted by muons crossing the water. The capability to detect muons, to reconstruct their tracks through  the scintillator was crucial to identify and tag cosmogenic contaminants (i.e. short living nuclei produced by muon spallation with scintillator components\,\cite{bib:Muons,bib:Cosmogenics}), especially $^{11}$C background. Muon tagging allows Borexino to also efficiently detect cosmogenic neutrons\,\cite{bib:Neutrons}, which occasionally are produced with high multiplicity, another crucial ingredient in $^{11}$C tagging.

\section*{Thermal insulation system and control}

The thermal stability of the Borexino detector is required to avoid undesired background variations due to the mixing of the scintillator inside the Inner Vessel. This mixing is caused by convective currents induced by temperature changes due to human activities in the underground Hall and to seasonal effects. A significant upgrade of the detector in this respect was carried out. 

Between May and December 2015, 900\,m$^2$ of thermal insulation was installed on the outside of the Borexino Water Tank.
In addition, the system used to recirculate water inside the WT was stopped in July 2015 to contribute to the inner detector thermal stability and allow its fluid to vertically stratify. 

The thermal insulation consists of two layers: an outer 10\,cm layer of Ultimate Tech Roll 2.0 mineral wool (thermal conductivity at 10$^\circ$C of 0.033 W/m/K) and an inner 10~cm layer of Ultimate Protect wired Mat 4.0 mineral wool reinforced with Al foil 65\,g/cm$^2$ with glass grid on one side (thermal conductivity at 10$^\circ$C of 0.030\,W/mK). The thermal insulation material is anchored to the WT with 20\,m long nails on a metal plate attached to the tank  (5\,nails/m$^2$). In addition, an active temperature control system (ATCS) was completed in January 2016. In ED Fig.~2 the Borexino WT is shown wrapped in thermal insulation.

A system of 66 probes with 0.07$^\circ$C resolution, the position of which is shown in ED Fig.~3, monitors the temperature of Borexino.
They are arranged as follows:
14 protruding 0.5\,m radially inward into the SSS (ReB probes) and in operation since October 2014, measure the temperature of the outer part of the buffer liquid (OB); 
14 mounted 0.5\,m radially outward from the SSS (ReW probes) and in operation since April 2015, measure the temperature of the water; 20 installed between the insulation layer and the external surface of the WT (WT probes) are in operation since May 2015; 4 located inside a pit underneath the Borexino WT are in operation since October 2015; 14 on the Borexino detector WT dome installed in early 2016. 
Since 2016 the average temperature of the floor underneath the detector in contact with the rock is 7.5$^\circ$C, while at the top of the detector it is 15.8$^\circ$C. 
This temperature difference corresponds to a naturally-driven gradient $\Delta T/\Delta z >0 \sim 0.5^\circ$C/m. Ensuring this gradient does not decrease it is the key to reducing convective currents, scintillator mixing, and consequently stabilizing the \po \,background for the CNO analysis.

Out of the last 14 probes, three are part of the Active Temperature Control System  (ATCS) kept in operation during the present data taking. The ATCS consists of a water based system made with copper tube coils installed on the upper part of the detector’s dome. The coils are in contact with the WT steel, with the addition of an Al layer to enhance the thermal coupling. A 3\,kW electric heater, a circulation pump, a temperature controller, and an expansion tank are connected to the coils. The ATCS trims the natural thermal gradient and is essential to eliminate convection motion. 

The Outer Detector head tank (a 70-liter vessel connected with the 1346\,m$^3$ volume of the SSS) is used as a sensitive detector thermometer. After the thermal insulation system installation the head tank had to be refilled with 289\,kg of PC because of the detector overall cooling and corresponding shrinkage. Calibration established the sensitivity of this thermometer to be of the order of 10$^{-2\,\circ}$C per 100\,mm change of fluid height. 

The deployment of both the thermal insulation and the temperature control systems were quickly effective in stabilizing the inner detector temperature. As of 2016 the heat loss due to the thermal insulation system was equal to 247\,W. Yet, changes of the experimental hall temperature induced residuals variations in the top buffer probes of the order of 0.3$^\circ$C/6 months. 
To further reduce these effects an active system to control the seasonal changes of the air temperature entering the experimental Hall and surrounding the Borexino WT was designed and installed in 2019. It consists of a 70\,kW electrical heater installed inside the inlet air duct, which has a capacity 12000\,m$^3$/h (in normal conditions). The heater is deployed just a few meters before the Hall main door. The temperature control is based on a master/slave architecture with a master PID controller that acts on a second slave PID controller. 
Probes deployed around the WT monitor the temperature of the air.
After commissioning, a set point temperature for the master PID of 14.5$^\circ$C is chosen. This system controls the temperature of the inlet air within approximately 0.05$^\circ$C. 

The thermal insulation, active temperature control of the detector, and the Hall C air temperature control have enabled remarkable temperature stability of the detector. ED Figure~4 shows the temperature time profile read by all probes since 2016. A stable temperature gradient was clearly established as needed to avoid mixing of the scintillator.

\section*{The Low Polonium Field and its properties}

After the completion of the thermal insulation (Phase-III), the Bismuth-210 background activity is measured from the \po \,activity assuming secular equilibrium of the $A=210$ chain. The measured \po \,rate is the sum of two contributions: a {\it scintillator} $^{210}$Po component supported by the $^{210}$Pb in the liquid ($^{210}$Po${^{\mathrm{S}}}$), which we assume to be stable in time and equal to the intrinsic rate of $^{210}$Bi in the scintillator, and a {\it vessel} component ($^{210}$Po${^{\mathrm{V}}}$). The latter has a 3D diffusive-like structure given by polonium
detaching from the Inner Vessel and migrating into the fiducial volume. The origin of this component is the \pb \,contamination of the vessel.
The \po \,migration process is driven by residual convective currents. A rough estimation of the migration length $\lambda_{\rm mig}$ obtained by fitting the spatial distribution of $^{210}$Po, is found to range between 50 and 100\,cm, which corresponds to a migration coefficient $D_{\rm mig}= (1.0 \pm 0.4) \times 10^{-9}$\,
m$^2$\,s$^{-1}$ (where we have used the relation $\lambda_{\rm mig} =\sqrt{D_{\rm mig}\tau_{\rm Po}}$ with the \po \,lifetime, $\tau_{\rm Po}$\,=\,199.7\,days).
This value is slightly lower than the diffusion coefficient $D_{\rm diff}\sim 1.5\times 10^{-9}$\,m$^2$\,s$^{-1}$ (corresponding to a diffusion length $\lambda_{\rm diff}\sim 20$\,cm), predicted by the Stokes-Einstein  formula\,\cite{bib:stokes} and observed for heavy atoms
in hydrocarbons\,\cite{bib:diff}. We interpret this difference as due to the presence of residual convective motions in Phase-III. These motions are localized in small regions and create a diffusive-like structure with an effective migration length $\lambda_{\rm mig} \gtrsim\lambda_{\rm diff}$.

The $\alpha$'s from \po \,decays are selected event-by-event with a highly efficient $\alpha/\beta$ pulse shape discrimination neural network method based on a \emph{Multi-layer Perceptron} (MLP)\,\cite{bib:tmva}. 
The resulting three-dimensional \po \,activity distribution, named the Low Polonium Field (LPoF), exhibits an effective migration profile with an almost stable minimum located above the detector equator (see 3D shape in ED Fig.~5, and dark blue regions in ED Fig.~6 top). The qualitative shape and approximate position of the LPoF is reproduced by fluid dynamical numerical simulations reported in\,\cite{bib:PoSimulation}.

 Assuming azimuthal symmetry around the detector $z$-axis, confirmed by 3D analysis, the \po minimum activity is determined by fitting LPoF with a 2D paraboloidal function: 
\begin{eqnarray} 
\label{eq:minpo}
\begin{split}
\frac{d^2R (^{210}\rm{Po})}{d(\rho^2) dz} = & \left[ R(^{210}{\rm Po}_{\rm min}) \epsilon_{\rm E}\epsilon_{\rm MLP} + R_\beta \right ] \times\\
&\times \left(1 +  \frac{\rho^2}{a^2} + \frac{(z-z_0)^2}{b^2} \right)\,,
\end{split}
\end{eqnarray}  
where $\rho^2 = x^2+y^2$, $a$ and $b$ are the paraboloid axes, $z_0$ is the position of the minimum along the $z$ axis, $\epsilon_{\rm E}$ and $\epsilon_{\rm MLP}$ are the efficiency of energy and MLP cuts used to select $\alpha$'s from $^{210}$Po decays, and $R_{\beta}$ is the residual rate of $\beta$ events after the selection of $\alpha$'s. 
The fit is initially performed in data bins of 2\,months, but compatible results are obtained using the bins of 1 month. ED Figure~6 top shows the result of the $z_0$ minimum position as a function of time. The minimum slowly moves along the $z$ direction by less than 20\,cm per month. In order to perform a better estimation of the \po minimum, we sum up all the time bins after aligning the 3D distributions with respect to $z_0$. Possible intrinsic biases, due to the minimum determination in different time intervals, have been minimized by blindly aligning the data from each time bin according to the $z_0$ inferred from the previous time interval.

The distribution of \po events after applying this procedure is shown in ED Fig.~6 bottom, where the LPoF structure is clearly visible. The final fit is then performed on 20\,tonnes of this aligned data set containing about 5000 \po \,events. From this fit we extract the \po \,minimum. This value might still have a small contribution from the vessel component (eq.\,\ref{eq:pomin}), i.e. $R(^{210}{\rm Bi})\,\leq\,R(^{210}{\rm Po}_{\rm{min}})$. Therefore this method provides only an upper limit for the \bi \,rate.
A companion analysis was performed using a 3D paraboloidal function. The 2D and 3D fits were performed with a standard binned likelihood and a Bayesian approach using non-informative priors. In particular, the latter was implemented with\,\textsc{MultiNest}\,\cite{bib:multinest1, bib:multinest2, bib:multinest3}, a nested sampling algorithm.

In addition, because the shape of the LPoF might show more complexity along the z-axis than a simple paraboloidal shape, a Bayesian framework was also used to perform the fit with a cubic spline along the $z$ axis. Splines are piecewise polynomials connected by \emph{knots}. The number of knots defines the complexity of the curve. To prevent over-fitting, a Bayesian factor analysis was used to decide on the most appropriate number of knots for the data set. While it was found that the splines were, in general, a better fit to the data (Bayes factor $> 10^2$), the final result is compatible to the simpler model within statistical uncertainties.
This result has been further cross-checked by fitting the \po \,distribution along different angular directions with a family of analytical functions found as solution of the Fick diffusion equation\,\cite{bib:fick} for the migration of decaying \po.
Possible biases have been quantified by testing the fit model on simulated LPoF patterns based on numerical fluid dynamical simulations. They were found to be negligible for our purpose.

\section*{Spatial uniformity and time stability of \bi}

The \bi \,independent constraint inferred from the LPoF can be extended over the whole fiducial volume if, and only if, the \bi \,itself is uniform in space.
Observation of the time stability of \bi \,rate, not strictly required if the time periods of the LPoF and main analyses are the same, can additionally crosscheck the overall robustness of the data set. 

We have evidence that at the beginning of Borexino Phase-II, after the purification campaign performed from 2010 to mid-2011, the \bi \,was not uniform: the cleanest part of the scintillator was concentrated on the top, partially out of the fiducial volume. In fact, the purification was performed in loop, taking the scintillator out from the bottom, purifying it, and re-inserting it from the top. For this reason, at the beginning of Phase-II the apparent \bi \,rate was  higher and slowly decreased in time as mixing was taking place, thanks to the strong pre-insulation convective currents.
 This decreasing trend stopped in early 2016 suggesting that the mixing had completed.
 Numerical fluid dynamical  simulations, performed using as input the velocity field obtained from  \po \,movements during the pre-insulation time,  confirm this hypothesis. 
 
  A more conservative approach, which uses heuristic arguments based on the effective migration of ions as measured from LPoF, suggests that \bi \,at the beginning of Phase-III (mid-2016) must be uniform at least within a volume scale of about 20\,m$^3$. This argument is also verified by means of fluid dynamics numerical simulations.

All the \emph{a priori} arguments and qualitative studies described above are confirmed \emph{a posteriori} by looking at the $\beta$ event rate  in optimized energy windows where the \bi \, signal-to-background ratio is maximal. The observed non-uniformity is then conservatively assigned only to \bi, contributing about 15\% to the overall rate in the selected energy window.

In order to test the spatial uniformity of the \bi \, rate in the fiducial volume and to associate a systematic uncertainty to its possible non-uniformity,
we split the spatial distribution into radial and angular components.

ED Figure~7 top shows the angular power spectrum of observed $\beta$ events (black points). The dark pink and pink bands are the allowed 1$\sigma$ and 2$\sigma$ regions respectively, obtained from $10^4$ Monte Carlo simulations of uniformly distributed events. The analysis is performed with the \textsc{HEALPix}\,\cite{bib:healpix} software package, available, \emph{e.g.},  for cosmic microwave background analysis. 

ED Figure~7 bottom shows the linear fit to the $r^3$ distribution of the $\beta$ events, expected to be flat for uniform spatial distribution, from which we determine the allowed residual non-uniformity along the radial direction. 

All these studies show no evidence for a sizeable non-uniformity of the $\beta$-like events, distribution inside the  fiducial volume. In particular, the rate measured in the LPoF is fully consistent with that measured in the total fiducial volume. 
This evidence further supports a very small systematic uncertainty on the \bi \,independent constraint. Combining in quadrature the uncertainties from the radial (0.52\,\cpd) and angular (0.59\,\cpd) components, we obtain a systematic error associated with the \bi \,spatial uniformity of 0.78\,\cpd.

Finally, we checked the \bi \,rate time stability applying two methods on the observed rate of $\beta$ events in the optimized energy windows:
(i) we studied the range of possible polynomial distortions; (ii) we performed a Lomb-Scargle spectral decomposition (see\,\cite{bib:seas} and references therein). We found no evidence of any relevant time variation besides the expected annual modulation due to solar neutrinos ($^7$Be-$\nu$'s contribute $>$60\% to the $\beta$ rate in the selected energy windows). Actually, the fact that we are able of seeing the tiny 3.3\% sinusoidal variation induced by the eccentricity of the Earth's orbit around the Sun, is in itself a further proof of the excellent time stability of the \bi \,rate. In particular, by studying the time dependence of the $\beta$-like events in the optimized window, the uncertainty on the \bi rate change is 0.18\,\cpd, which is indeed negligible as compared with the global error quoted in Eq.\,\ref{eq:penalty}.

We note that, even after complete mixing, the true \bi \,rate is not perfectly constant in time, as it must follow the decay rate of the parent \pb ($\tau = 32.7$\,y). This effect is not detectable over the $\sim$3\,years time period of our analysis, but for substantially longer periods
it could be used for better constraining the \bi \,by fitting its long-lived temporal trend.

\section*{Details of the CNO analysis}

The analysis presented in this work is based on the data collected from June 2016 to February 2020 (Borexino Phase-III) and is performed in a fiducial volume (FV) defined as $r$\,$<$\,2.8\,m and $-1.8$\,m$<$\,$z$\,$<$\,2.2\,m ($r$ and $z$ being the reconstructed radial and vertical position, respectively). The total exposure of this dataset corresponds to 1072\,days $\times$ 71.3\,tonnes.

In Borexino, the energy of each event is given by the number of collected photoelectrons, while its position is determined by the photon arrival times at the PMTs. 
The energy and spatial resolution in Borexino has slowly deteriorated over time due to the steady loss of PMTs (the average number of active channels in Phase-III is 1238) and it is currently $\sigma_E/E \approx$6\% and $\sigma_{x,y,z} \approx$11\,cm for 1\,MeV events in the center of the detector.

Events are selected by a sequence of cuts, which are specifically designed to veto muons and cosmogenic isotopes, to remove $^{214}$Bi\,-\,$^{214}$Po fast coincidence events from the $^{238}$U chain, electronic noise, and external background events. The fraction of neutrino events lost by this selection criteria is measured with calibration data to be of the order of 0.1\% and is therefore negligible. More details on data selection can be found in\,\cite{bib:Nusol}.

The main backgrounds surviving the cuts and affecting the CNO analysis are: \bi and $^{210}$Po in secular equilibrium with $^{210}$Pb which, as discussed thoroughly in the previous paragraphs, have a rate in Borexino Phase-III of $\leq$\,(11.5\,$\pm$\,1.3)\,\cpd;  $^{210}$Po$^{\rm V}$ from the vessel;  $^{85}$Kr ($\beta$, Q-value\,=\,687\,keV); 
$^{40}$K ($\beta$ and $\gamma$, Q-value\,=\,1460\,keV), 
$^{11}$C ($\beta^+$, Q-value\,=\,960\,keV; $\tau$\,=\,30\,min), 
which is continuously produced by cosmic muons crossing the scintillator; $\gamma$ rays emitted by $^{214}$Bi, $^{208}$Tl, and $^{40}$K from materials external to the scintillator (buffer liquid, PMTs, stainless steel sphere, etc.).

CNO neutrinos are disentangled from residual backgrounds through a multivariate analysis, which includes the energy and radial distributions of the events surviving the selection. Data are split into two complementary data sets: the {\it TFC-subtracted} spectrum, where $^{11}$C is selectively filtered out using the muon-neutron-positron three-fold coincidence algorithm (TFC)\,\cite{bib:pep, bib:be7Long} and the {\it TFC-tagged spectrum}, enriched in $^{11}$C.
The TFC is a space and time coincidence vetoing the $^{11}$C $\beta^+$ decay events, by tagging the spallation muon and the neutron capture from the reactions: $\mu + ^{12}{\rm C} \rightarrow ^{11}{\rm C} + n$ and $n + p \rightarrow d + \gamma$.
The reference shapes, \emph{i.e.} the probability distribution functions (PDFs) for signal and backgrounds used in the fit, are obtained through a complete \textsc{Geant4}-based Monte Carlo code\,\cite{bib:MCPaper}, which simulates all physics processes occurring in the scintillator, including energy deposition, photon emission, propagation, and detection, generation and processing of the electronic signal. The simulation takes into account the evolution in time of the detector response and produces data that are reconstructed and selected following the same pipeline of real data. The relevant input parameters of the simulation, mainly related to the optical properties of the scintillator and of the surrounding materials, have been initially obtained through small-scale laboratory tests and subsequently fine-tuned on calibration data, reaching an agreement at the sub-percent level\,\cite{bib:bxcalib}. 
Data is then fitted as the sum of signal and background PDFs: the weights of this sum (the energy integral of the rates with zero threshold of each component in Borexino) are the only free parameters of the fit. 
The details of the multivariate fit tool, used also to perform other solar neutrino analysis in Borexino, are described thoroughly in\,\cite{bib:Nature-2018} and\,\cite{bib:Nusol}. Differently from the previous comprehensive {\it pp} chain analysis  the fit is performed performed between 320 and 2640\,keV, thus excluding the contribution of $^{14}$C decays and its pile-up. This choice is motivated by the loss of energy and position resolutions due to the decreased number of active channels in Phase-III, which has an impact mainly in the low energy region. 

In addition to the energy shape, other information is exploited to help the fit to disentangle the signal from background:
the $^{11}$C $\beta^+$ events are tagged by TFC, and contributions from the external backgrounds ($^{208}$Tl, $^{214}$Bi, and $^{40}$K) are further constrained thanks to their radial distribution.

In order to enhance the sensitivity to CNO neutrinos, the $pep$ neutrino rate is constrained to the value ($2.74 \pm 0.04$)\,cpd per 100\,t derived from  a global fit\,\cite{bib:Vissani2019, bib:bergstrom} to solar neutrino data and imposing the $pp/pep$ ratio and the solar luminosity constraint, considering the MSW matter effect on the neutrino propagation, as well as the errors on the neutrino oscillation parameters. As discussed in the main text, the spectral fit has little capability to disentangle events due to CNO neutrino interactions and \bi \,decay.
Therefore, we use the results of the independent analysis on the $^{210}$Po distribution in the LPoF to set an upper limit to the \bi rate of (11.5\,$\pm$\,1.3)\,\cpd.

 The results of the simultaneous multivariate fit are given in ED Fig.~8, showing the TFC-subtracted and TFC-tagged energy spectra, and in ED Fig.~9, demonstrating the fit of the radial distribution.
 The fit is performed in the energy estimator $\mathcal{N}_{\rm hits}$ (defined as the sum of all photons triggering a PMT, normalized to 2000 active PMTs) and the results are reported also in keV. 
The $p$-value of the fit is 0.3, demonstrating fair agreement between data and the underlying fit model. The fit clearly prefers a non-zero CNO neutrino rate as shown in the log-likelihood profile of Fig.~4 (dashed black curve).

 Many sources of possible systematic errors have been considered.
 
 The systematic error associated to the fit procedure has been studied by performing 2500 fits with slightly changed conditions (different fit ranges and binning) and was found to be negligible with respect to the statistical uncertainty.
 
 Since the multivariate analysis relies critically on the simulated PDFs of signal and backgrounds, any mismatch between the realistic and simulated energy shapes can alter the result of the fit and bias the significance on the CNO neutrinos. In order to study the impact of these possible mismatches, we simulated over a million of pseudo-data sets with the same exposure of Phase-III, injecting deformations in the signal and background shapes, following\,\cite{bib:cousin}. Each data-set is then fitted  with the standard non-deformed PDFs. The study was performed injecting different values of CNO including the one obtained by our best fit. 
 We studied the impact of the following sources of deformations:
 \begin{itemize}
\item Energy response function: inaccuracies in the energy scale (at the level of $\simeq$\,0.23\%) and in the description of non-uniformity and non-linearity of the response (at the level of $\simeq$\,0.28\% and $\simeq$\,0.4\%, respectively). The size of the applied deformations has been chosen in the range allowed by calibration data and  by data from specific internal backgrounds ($^{11}$C and $^{210}$Po) taken as reference ``standard candles";
 \item Deformations of the $^{11}$C spectral shape induced by cuts to remove noise events, not fully taken into account by the Monte Carlo PDFs (at the level of 2.3\%);
\item Spectral shape of $^{210}$Bi: we have studied the systematic error associated to the shape of the forbidden $\beta$-decay of $^{210}$Bi simulating data with alternative spectra (found in\,\cite{bib:Carles1996} and in\,\cite{bib:210BiDerbin}) with respect to the default one\,\cite{bib:Daniel1962}. Differences in the shapes may be as large as 18\%;
\end{itemize}
From this Monte Carlo study we evaluate the CNO systematic error due to  a mismatch between real and simulated PDFs to be \FinalSys\,\cpd. This uncertainty is deduced by comparing the CNO output distributions from toy Monte Carlo's with and without injecting systematic distortions as described above.

 In order to evaluate the significance of our result in rejecting the no-CNO
 hypothesis, we performed a frequentist hypothesis test using a profile
 likelihood test statistics $q$ defined following Ref.\,\cite{bib:cowan} as: 
 \begin{equation}
 \label{eq:q}
 q=-2 \log  \frac{\mathcal{L} (\rm {CNO=0})}{\mathcal{L} (\rm {CNO})},
 \end{equation}
 where $\mathcal{L}(\rm {CNO=0})$ and $\mathcal{L}(\rm {CNO})$ is the maximum likelihood obtained by keeping the CNO rate fixed to zero or free, respectively.
 ED Figure~10, shows the $q$ distribution obtained from 13.8 millions of pseudo-data sets simulated with deformed PDFs (see discussion above) and no-CNO injected ($q_0$, grey  curve). In the same plot, the theoretical $q_0$ distribution in case of no PDF deformation is shown (blue curve). The result on data obtained from the fit is the black line ($q_{\rm data}=$\,30.05). 
 
 The plot in ED Fig. 10, allows us to reject the CNO\,=\,0 hypothesis with a significance better than \FinalSigma  $\sigma$ at 99.0\% C.L.\,\cite{bib:jeffery}.
 This construction is consistent with the significance evaluation of $5.1\sigma$, reported in the main text, by means of the quantiles of the profile likelihood folded with the systematic uncertainty.

 In ED Fig.~10, we also provide as reference the $q$ distribution (red) obtained with one million pseudo-data sets including systematic deformations and injected CNO rate equal to $7.2$ cpd per 100 t, i.e., our best fit value.

 A cross-check of the main analysis has been performed with a nearly independent method ({\it counting analysis}), in which we simply count events in an optimized energy window (region of interest, ROI) and subtract the contributions due to known backgrounds in order to reveal the CNO signal. This method is simpler, albeit less powerful, with respect to the multivariate fit and is less prone to possible correlations between different species. However, while the multivariate analysis implicitly checks the validity of the background model by the goodness of the fit, the counting analysis relies completely on the assumption that there are no unknown backgrounds contributing to the ROI. 
 
 The counting analysis is based on a different energy estimator than the multivariate analysis ($\mathcal{N}_{\rm pe}$, the total charge of all hits, normalized to 2000 active channels) and relies on a different response function (analytically derived, instead of Monte Carlo based) to determine the percentage of events for each of the signal and background species falling inside the ROI. The chosen ROI, (780\,-\,885)\,keV, is obtained optimizing the CNO signal-to-background ratio. An advantage of this method is that in the ROI some of the backgrounds which affect the multivariate analysis (like $^{85}$Kr and $^{210}$Po) are not present or contribute less than 2\% (\emph{e.g.} external backgrounds). The  count rate is dominated by CNO, {\it pep}, and \bi (80\%), with smaller contributions from $^7$Be neutrinos and residual $^{11}$C (18\%).
 The rate of {\it pep} neutrinos and \bi \,are constrained to the same values used in the multivariate fit. Note that while in the spectral fit the \bi \,rate is left free to vary between 0 up to (11.5\,$\pm$\,1.3)\,\cpd  (the upper limit determined in the LPoF analysis), the counting analysis conservatively constrains it to the maximum value with a Gaussian error of 1.3\,\cpd. The $^7$Be neutrino rate is sampled uniformly between the LZ (43.7 $\pm$ 2.5 \cpd) and the HZ (47.9 $\pm$ 2.8 \cpd) values predicted by the Standard Solar Model\,\cite{bib:LowZHighZ} with 1$\sigma$ error, while the $^{11}$C rate is obtained from the average Borexino Phase-II results with an additional conservative error of 10\% deriving from uncertainties on the energy scale (quenching of the 1\,MeV annihilation $\gamma$'s).
 The CNO rate is obtained by subtracting all background contributions defined above and by propagating the uncertainties by randomly sampling their rates from Gaussian distributions with proper widths. Note that the uncertainty related to the energy response (which affects the percentage of the spectrum of each component falling in the ROI) also contributes to the total error associated to the count rate of each species.
 
 The CNO rate obtained with this method is demonstrated by the red histogram in Fig.~4. The mean value and width of the distribution are (5.6\,$\pm$\,1.6)\,\cpd, confirming the presence of CNO at the 3.5$\sigma$ level.

The counting analysis shows that the core of the sensitivity to CNO neutrinos in Borexino  mainly comes, as expected, from a narrow energy region in which the contributions from CNO, \emph{pep}, $^{210}$Bi are dominant over the residual backgrounds, as discussed in\,\cite{bib:sensitivity-paper}. The multivariate fit, on the other hand, effectively exploits additional information contained in the data with a substantial enhancement of the CNO solar neutrino signal significance.

\end{methods}

\textsf{
\noindent
{\bf FIGURE 1 $|$\,\,\,CNO nuclear fusion sequences and the energy spectra of solar neutrinos.} \emph{Upper plot}: the double \cno cycle in the Sun, where sub-cycle I is dominant. The colored arrows indicate the reaction rates integrated over the Sun's volume. The rate of $^{17}$O($\alpha$,\,$p$)$^{14}$N reaction is below the low end of the color scale (dashed arrow). \emph{Lower plot}: energy spectra of solar neutrinos from the $pp$ chain (grey, $pp$, $pep$, $^7$Be, $^8$B, and $hep$) and CNO cycle (in colour). The two dotted lines indicate electron capture\,\cite{Bahcall1990, Stonehill2004, Villante2015}. For mono-energetic lines the flux is given in cm$^{-2}$\,s$^{-1}$. 
\\
\\
\noindent
{\bf FIGURE 2 $|$\,\,\,Spectral fit of the Borexino data.} Distribution of the electron recoil energy scattered by solar neutrinos in Borexino (black points) and corresponding spectral fit (magenta). 
CNO-$\nu$, \bi, and \emph{pep}-$\nu$ are highlighted in solid red, dashed blue, and dotted green, respectively. All other components are in grey. The yellow band represents the region with
the largest signal-to-background ratio for CNO-$\nu$.
\\
\\
\noindent
{\bf FIGURE 3 $|$\,\,\,Space and time distribution of the \po\,activity.} \po rate in Borexino in \cpd (rainbow color scale) as a function of time in small cubes of about 3\,tonnes each ordered from the bottom, ``0", to the top, ``58", along the vertical direction (Latest update: May 2020). All cubes are selected inside a sphere of radius $r = 3$\,m. The red curve with its red scale on the right represents the average temperature in the innermost region surrounding the nylon vessel.
The dashed vertical lines indicate the most important milestones of the temperature stabilisation program: 1. Beginning of the ``Insulation Program''; 2. Turning off of the water recirculation system in the Water Tank; 3. First operation of the active temperature control system; 4. Change of the active control set point; 5. Installation and commissioning of the Hall C temperature control system. The white vertical bands represent different DAQ interruptions due to technical issues.
\\
\\
\noindent
{\bf FIGURE 4 $|$\,\,\,Results of the CNO counting and spectral analyses.} \emph{Left.} {\it Counting analysis bar chart}. The height represents the number of events allowed by the data for CNO-$\nu$ and backgrounds in ROI; on the left, the CNO signal is minimum and backgrounds are maximum, while on the right,  CNO is maximum and backgrounds are minimum. It is clear from this figure that CNO cannot be zero. \emph{Right.} CNO-$\nu$ rate negative log-likelihood profile directly from the multivariate fit (dashed black line) and after folding in the systematic uncertainties (black solid line). Histogram in red:
 CNO-$\nu$ rate obtained from the counting analysis. Finally, the blue, violet, and grey vertical bands show 68\% confidence intervals (C.I.) for the SSM-LZ ($3.52 \pm 0.52$ \cpd) and SSM-HZ ($4.92 \pm 0.78$ \cpd)\,\cite{Vinyoles:2016djt,bib:sensitivity-paper} predictions and the Borexino result (corresponding to black solid-line log-likelihood profile), respectively.
\\
\\
\noindent
{\bf ED FIGURE 1 $|$\,\,\,The Borexino detector.} Schematic view of the structure of the Borexino apparatus; from inside to outside: the liquid scintillator, the buffer liquid, the stainless steel sphere with the photomultipliers, and the water tank.
\\
\noindent
{\bf ED FIGURE 2 $|$\,\,\,The Borexino detector after the thermal stabilisation.} The Borexino Water Tank after the completion of the thermal insulation and the active temperature control system deployment.
\\
\\
\noindent
{\bf ED FIGURE 3 $|$\,\,\,Temperature probes of the Borexino detector.} Distribution of temperature probes around and inside the Borexino detector. For simplicity, the probes on the WT dome and in the pit below the detector are not shown.
\\
\noindent
{\bf ED FIGURE 4 $|$\,\,\,Temperature time evolution in Borexino.} Temperature as a function of time in different volumes of the Borexino detector. The vertical dashed lines show: the activation of the temperature control system on the dome of the Water Tank, the set-point change, and the activation of the air control system in the experimental hall.
\\
\\
\noindent
{\bf ED FIGURE 5 $|$\,\,\,The Low Polonium Field in the Borexino scintillator.} Three-dimensional view of the \po \,activity inside the entire nylon vessel (see colour code). The innermost blue region contains the LPoF (black grid). The white grid is the software-defined fiducial volume.
\\
\\
\noindent
{\bf ED FIGURE 6 $|$\,\,\,Analysis of the Low Polonium Field.} \emph{Top.} The rate of \po \,in cylinders of 3\,m radius and 10\,cm height located along the $z$ axis from -2\,m to 2\,m, as a function of time with 1 month binning. The dashed lines indicate the $z$ coordinate of the fiducial volume. The markers show the positions of the center of the LPoF  obtained with two fit methods: \emph{paraboloid} (red) and \emph{spline} (white). Both fit methods follow the dark blue minimum of the \po \,activity well. 
    The structure visible in mid-2019 is due to a local instability produced by a tuning of the active temperature control system. This transient has no impact on the final result.
\emph{Bottom.} Distribution of \po \,events after the blind alignment of data using the $z_0$ from the paraboloidal fit (red markers in ED Fig.~6 \emph{top}). The red solid lines indicate the paraboloidal fit within 20\,tonnes with Eq.\,\ref{eq:minpo}. 
\\
\\
\noindent
{\bf ED FIGURE 7 $|$\,\,\,Angular and radial uniformity of the $\beta$ events in the optimized energy window.} \emph{Top.} Angular power spectrum as a function of the multipole moment $l$ of observed $\beta$ events (black points) compared with 10$^4$ uniformly distributed events from Monte Carlo simulations at one (dark pink) and two $\sigma$ C.L. (pink). Data are compatible with a uniform distribution within the uncertainty of 0.59 \cpd. \emph{Inset}: Angular distribution of the $\beta$ events.
\emph{Bottom.} Normalized radial distribution of $\beta$ events $r/r_0$ (black points), where $r_0$\,=\,2.5\,m is the radius of the sphere surrounding the analysis fiducial volume. The linear fit of the data (red solid line) is shown along with the 1$\sigma$ (yellow) and 2$\sigma$ (green) C.L. bands.
    The data are compatible with a uniform distribution within 0.52\,\cpd.
\\
\\
\noindent
{\bf ED FIGURE 8 $|$\,\,\,Multivariate fit of the Borexino data: energy distributions.} Full multivariate fit results for the TFC-subtracted (\emph{left}) and the TFC-tagged (\emph{right}) energy spectra with corresponding residuals. In both figures the magenta lines represent the resulting fit function, the red line is the CNO neutrino electron recoil spectrum, the green dotted line is the \emph{pep} neutrino electron recoil spectrum, the dashed blue line is the \bi beta spectrum, and in grey we report the remaining background contributions. 
\\
\\
\noindent
{\bf ED FIGURE 9 $|$\,\,\,Multivariate fit of the Borexino data: radial distribution.} Radial distribution of events in the multivariate fit. The red line is the resulting fit, the green line represents the internal uniform contribution and the blue line shows the non-uniform contribution from the external background.
\\
\\
\noindent
{\bf ED FIGURE 10 $|$\,\,\,Frequentist hypothesis test for the CNO observation.} Distribution of the test statistics $q$ (eq.\,\ref{eq:q}) from  Monte Carlo pseudo-data sets. The grey distribution $q_0$ is obtained with no CNO simulated data and includes the systematic uncertainty. The black vertical line represents $q_{\rm data}=30.05$. The corresponding $p$-value of $q_0$ with respect to $q_{\rm data}$ gives the significance of the CNO discovery ($>$\FinalSigma$\sigma$ at 99\% C.L.). For comparison, in blue is the $q_0$ without the systematics. The red histogram represents the expected test statistics distribution for injected CNO rate equal to $7.2$ cpd per 100\,t, i.e. our best fit value. 
}

\begin{figure}
\centering
\includegraphics[width=0.75\textwidth]{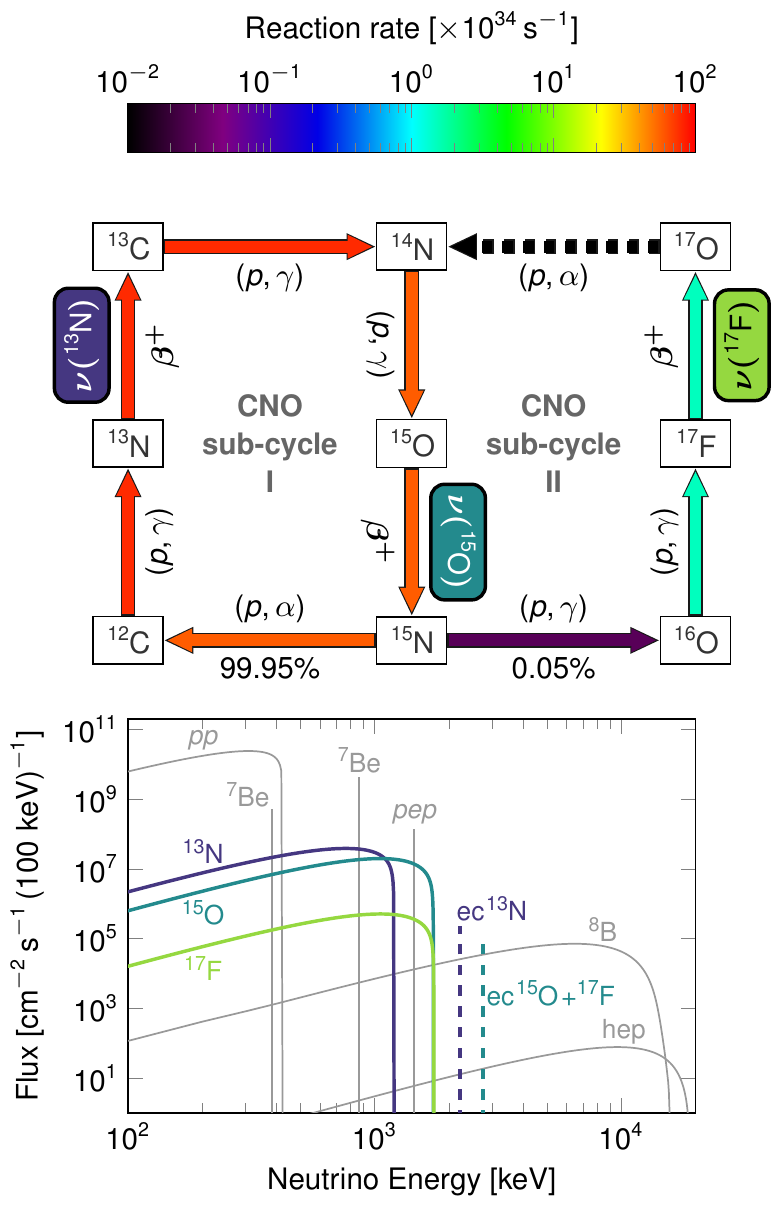}
\caption{\textsc{Figure-1}}
\end{figure}

\begin{figure}
\centering
\includegraphics[width=\textwidth]{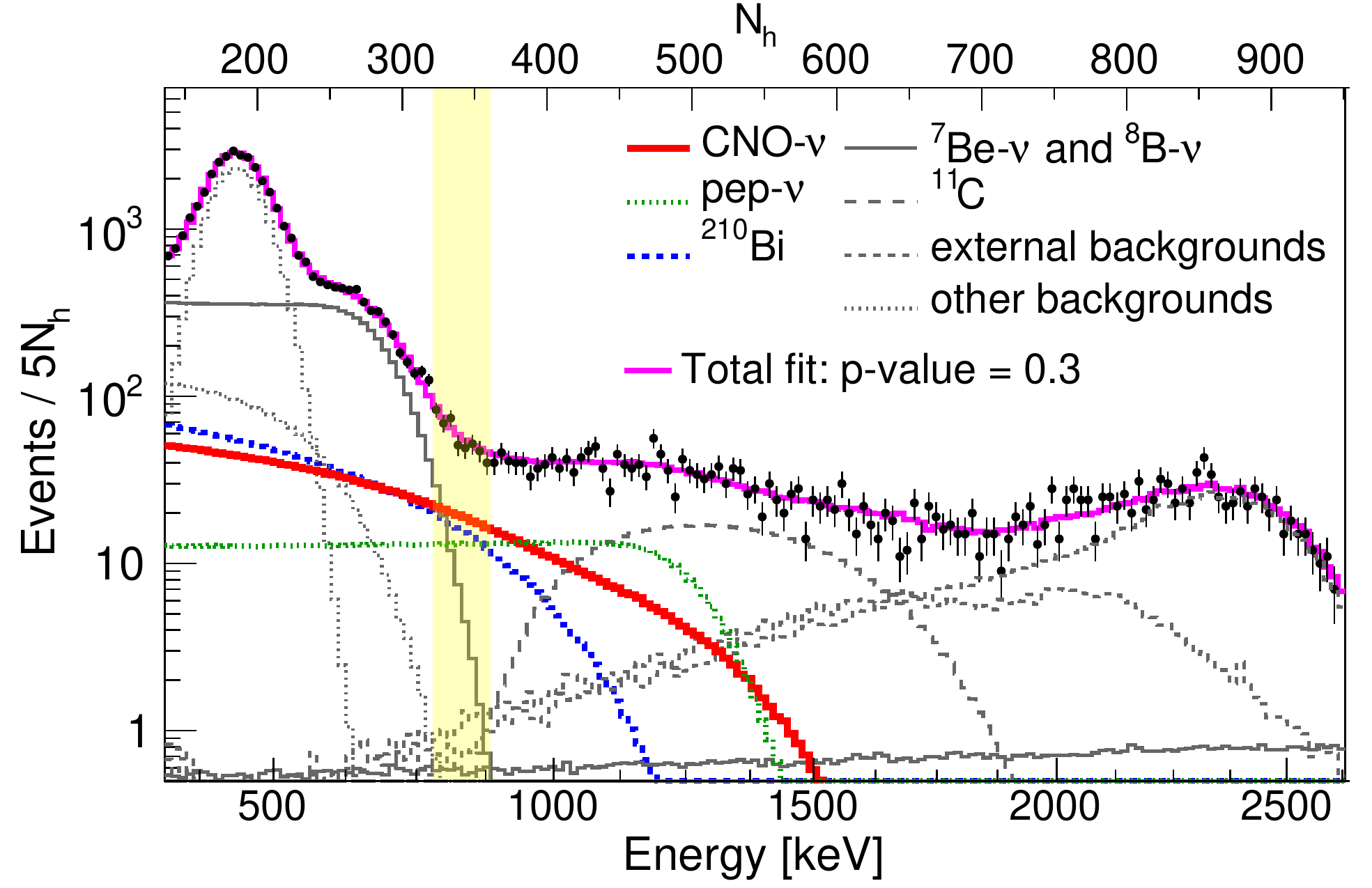}
\caption{\textsc{Figure-2}}
\end{figure}

\begin{figure}
\centering
\includegraphics[width=\textwidth]{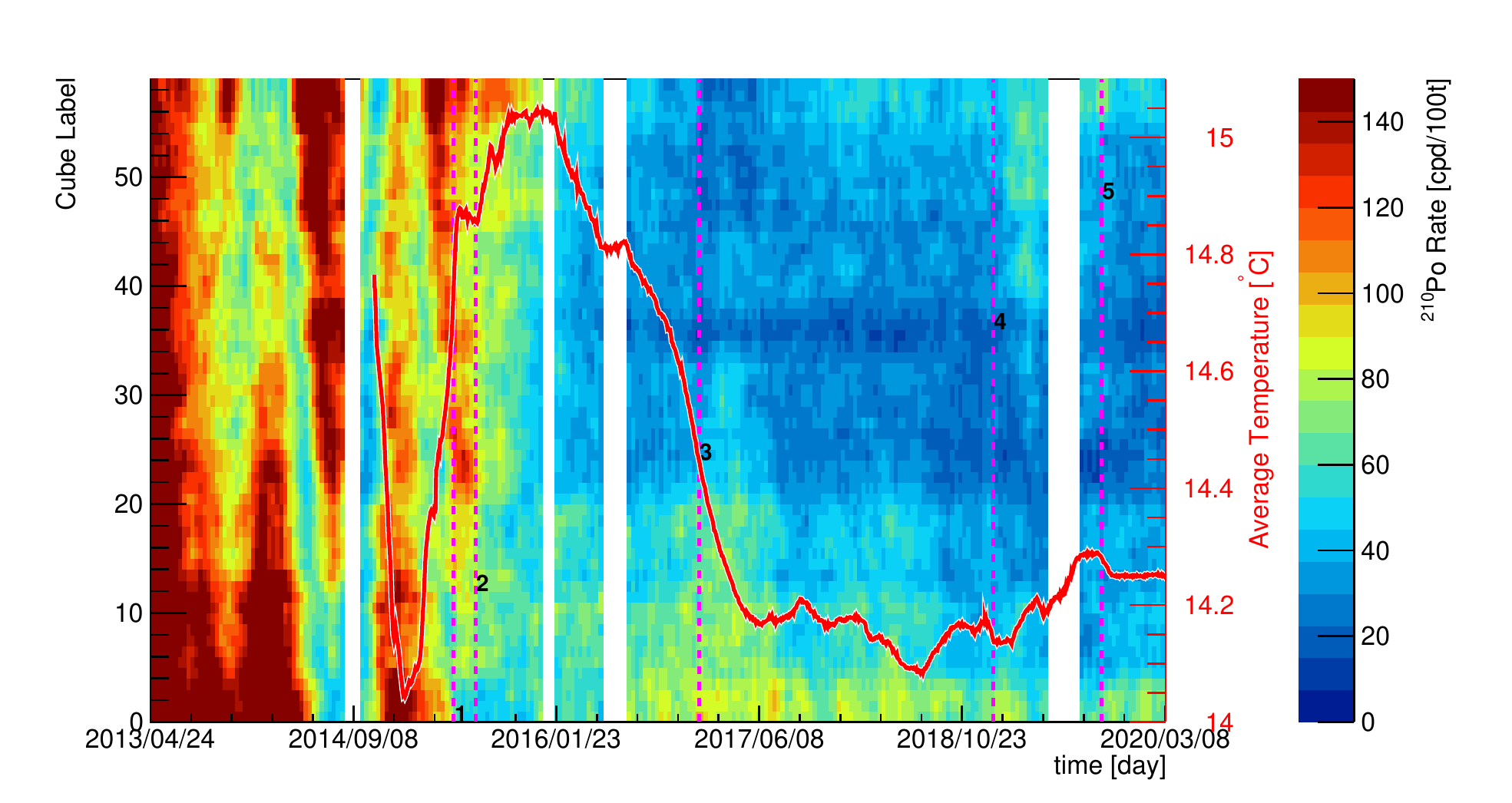}
\caption{\textsc{Figure-3}}
\end{figure}

\begin{figure}
\centering
\includegraphics[width=\textwidth]{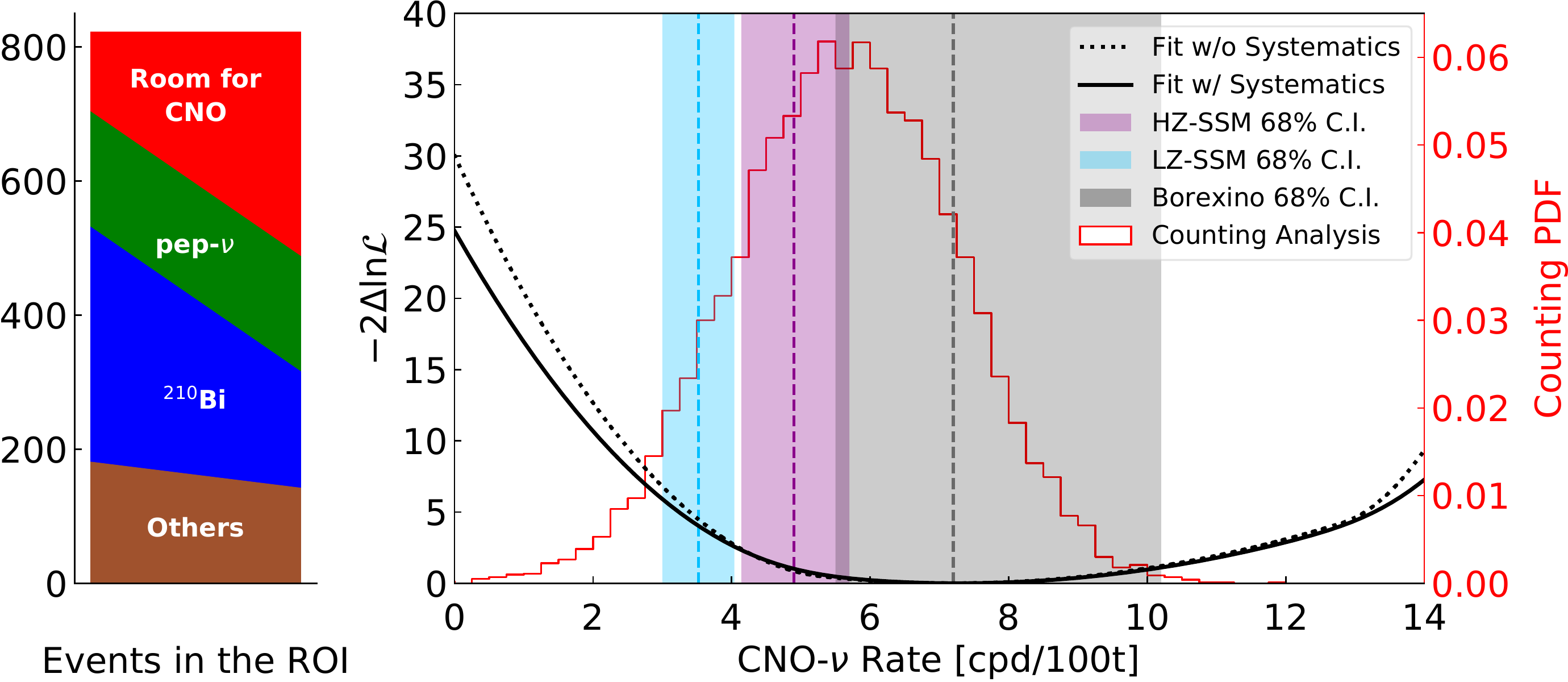}
\caption{\textsc{Figure-4}}
\end{figure}


\begin{figure}
\centering
\includegraphics[width=\textwidth]{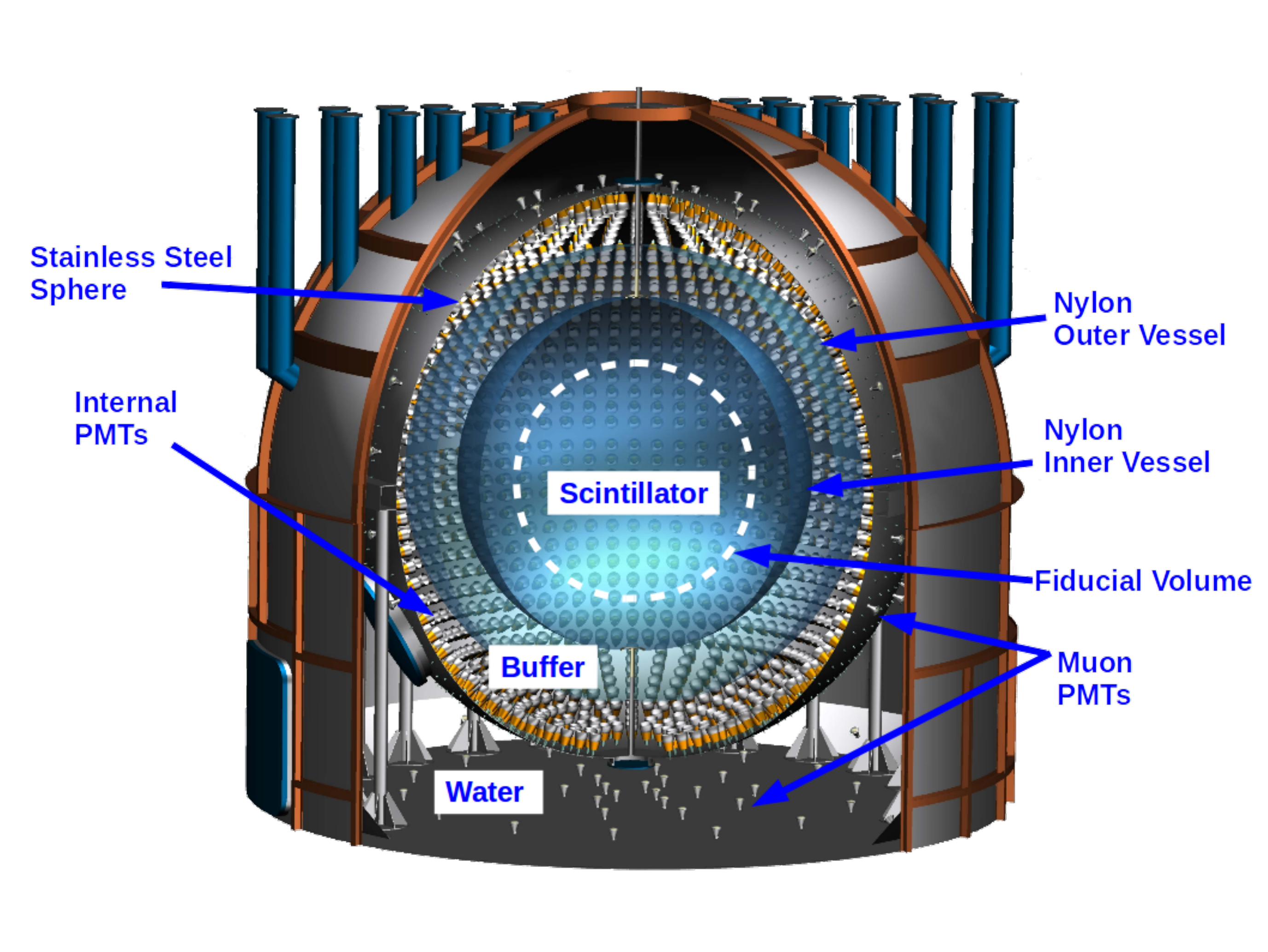}
\caption{\textsc{ED-Figure-1}}
\end{figure}

\begin{figure}
\centering
\includegraphics[width=\textwidth]{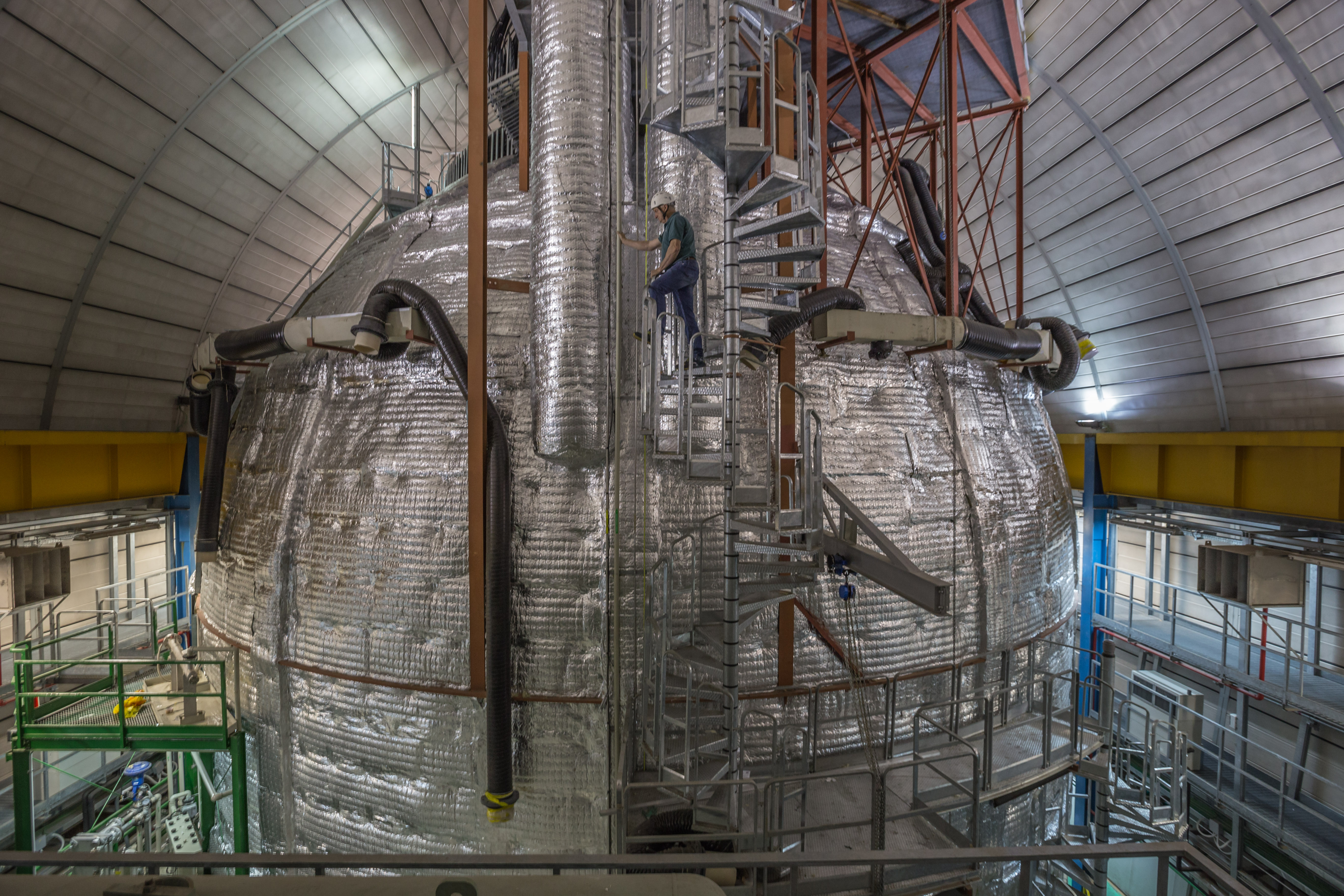}
\caption{\textsc{ED-Figure-2}}
\end{figure}

\begin{figure}
\centering
\includegraphics[width=\textwidth]{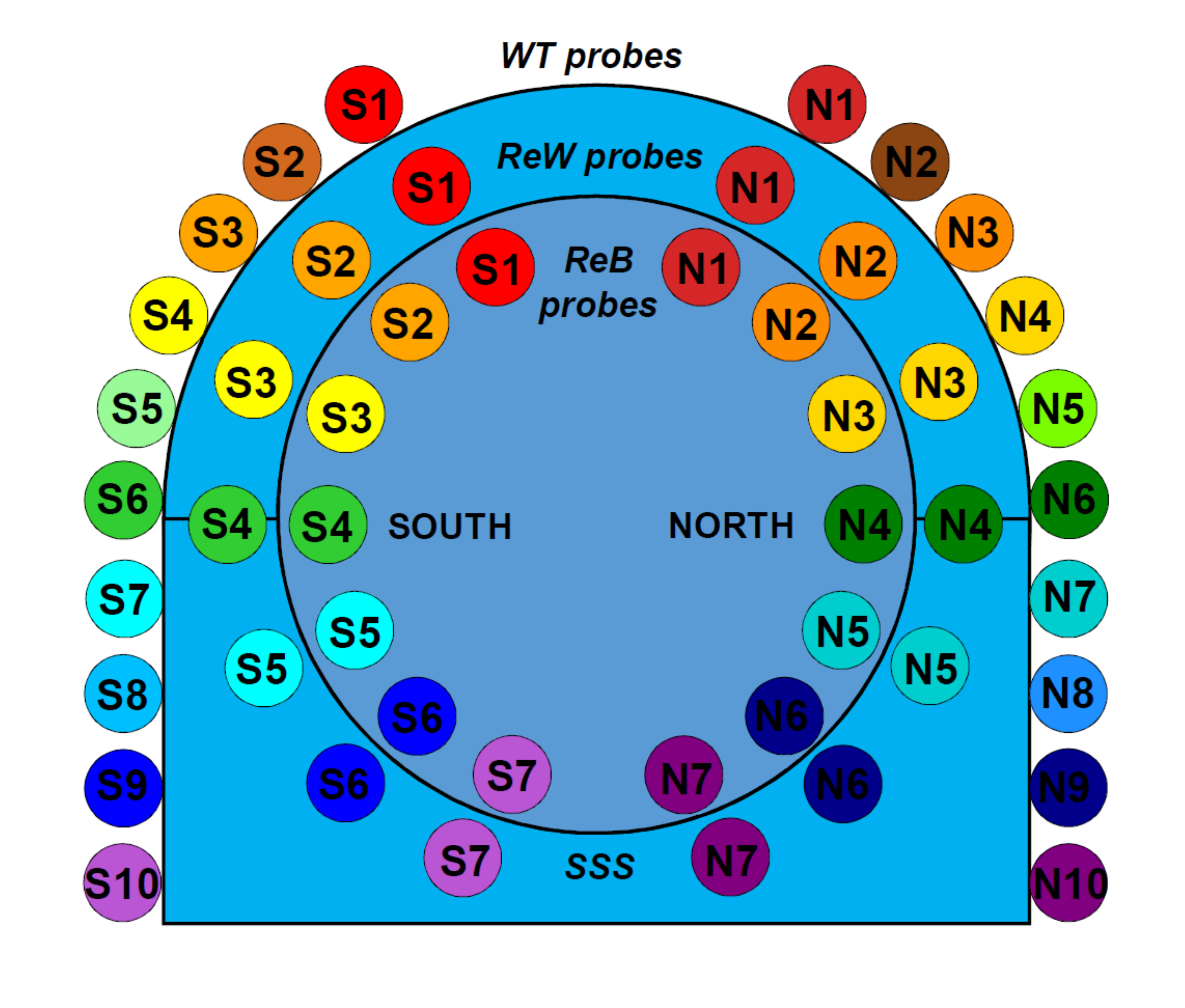}
\caption{\textsc{ED-Figure-3}}
\end{figure}

\begin{figure}
\centering
\includegraphics[width=\textwidth]{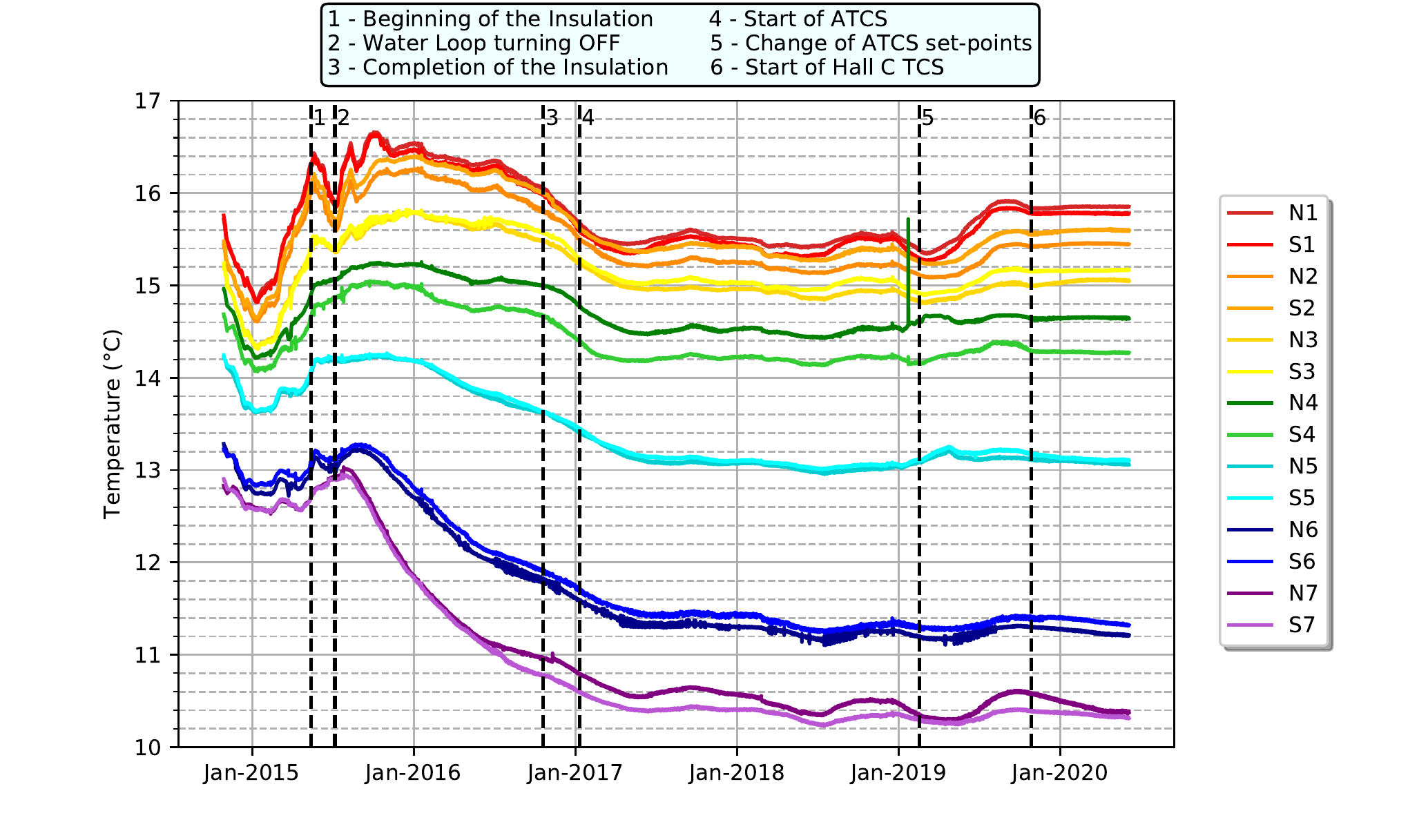}
\caption{\textsc{ED-Figure-4}}
\end{figure}

\begin{figure}
\centering
\includegraphics[width=\textwidth]{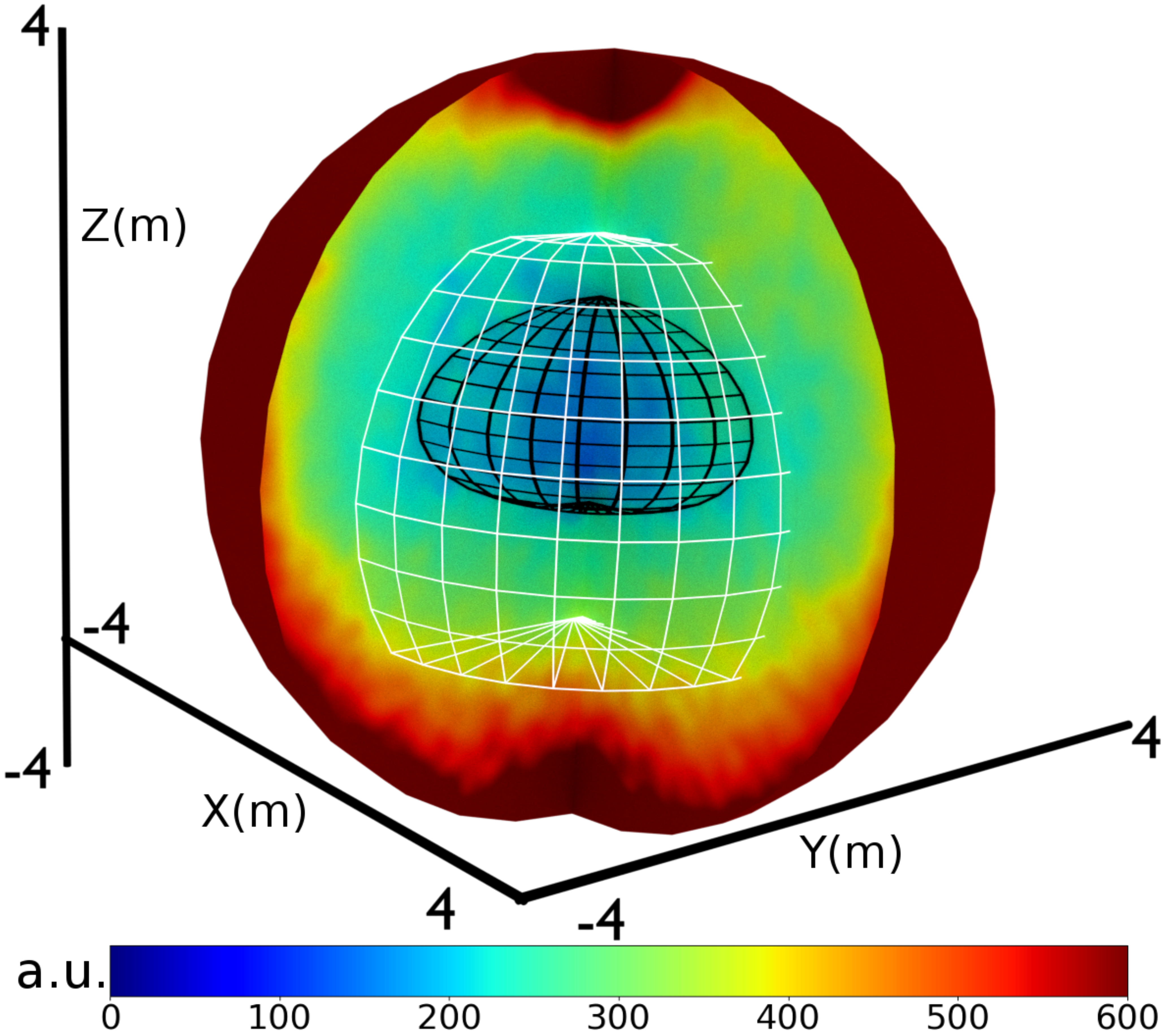}
\caption{\textsc{ED-Figure-5}}
\end{figure}

\begin{figure}
\centering
\includegraphics[width=\textwidth]{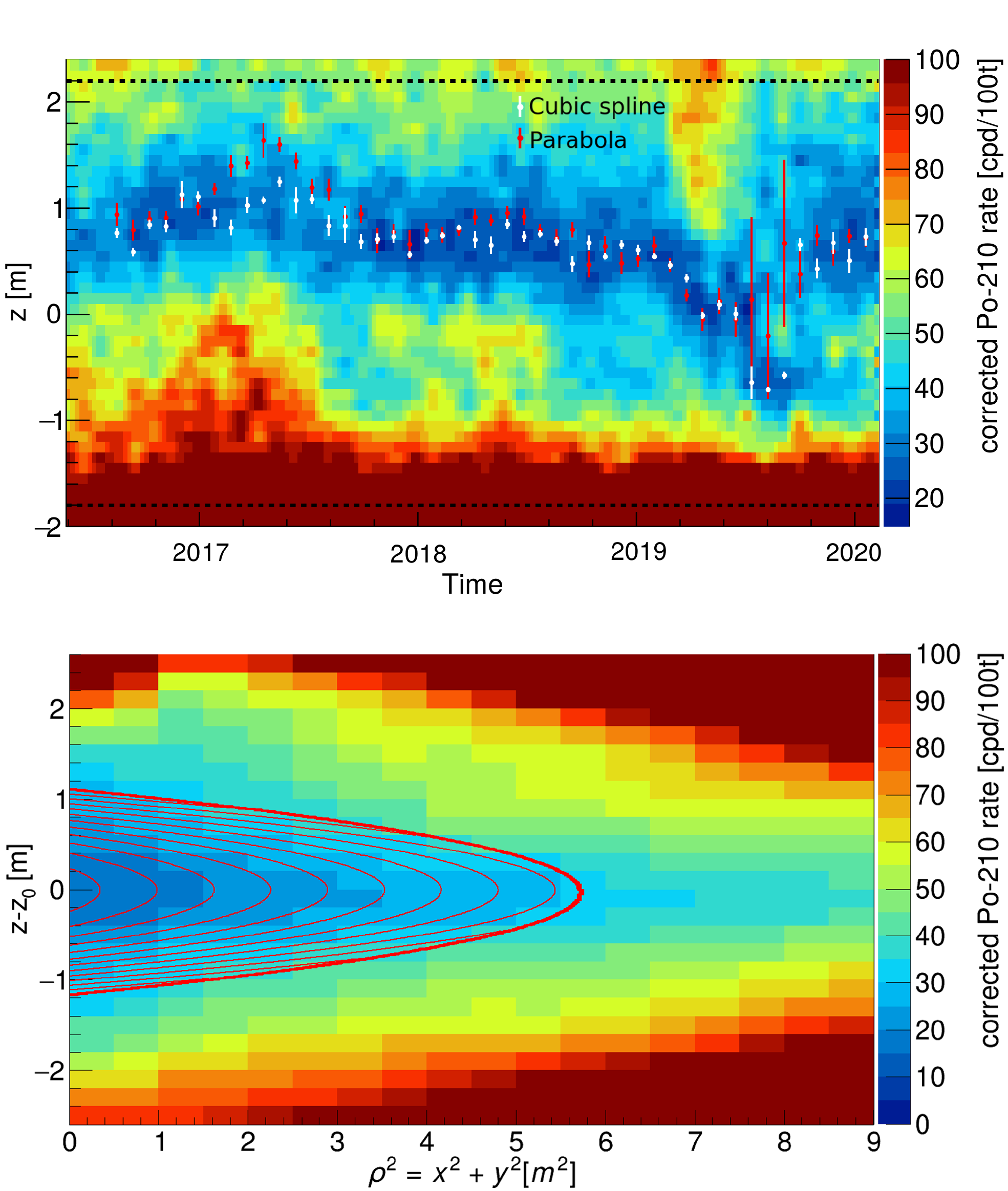}
\caption{\textsc{ED-Figure-6}}
\end{figure}

\begin{figure}
\centering
\includegraphics[width=\textwidth]{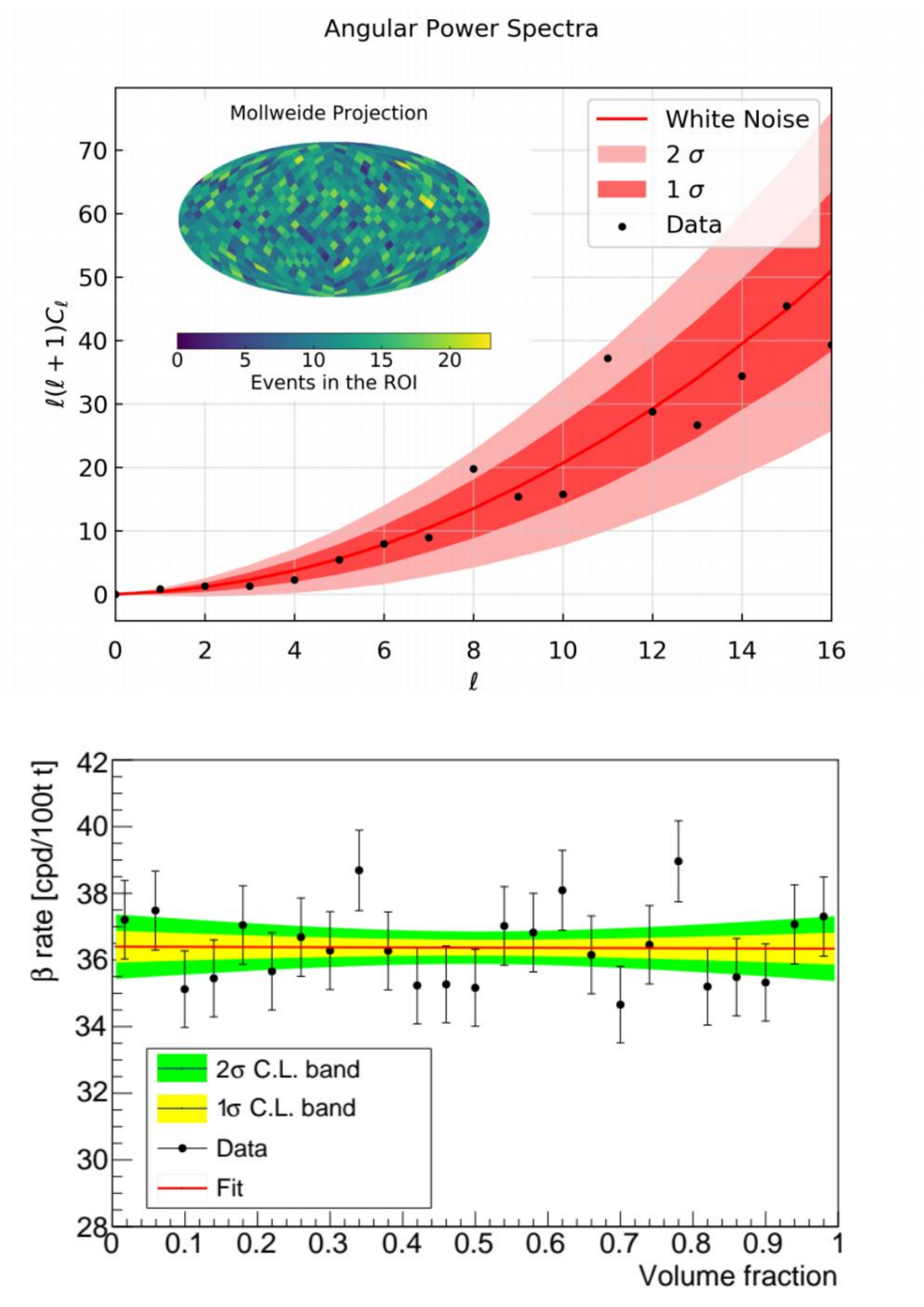}
\caption{\textsc{ED-Figure-7}}
\end{figure}

\begin{figure}
\centering
\includegraphics[width=\textwidth]{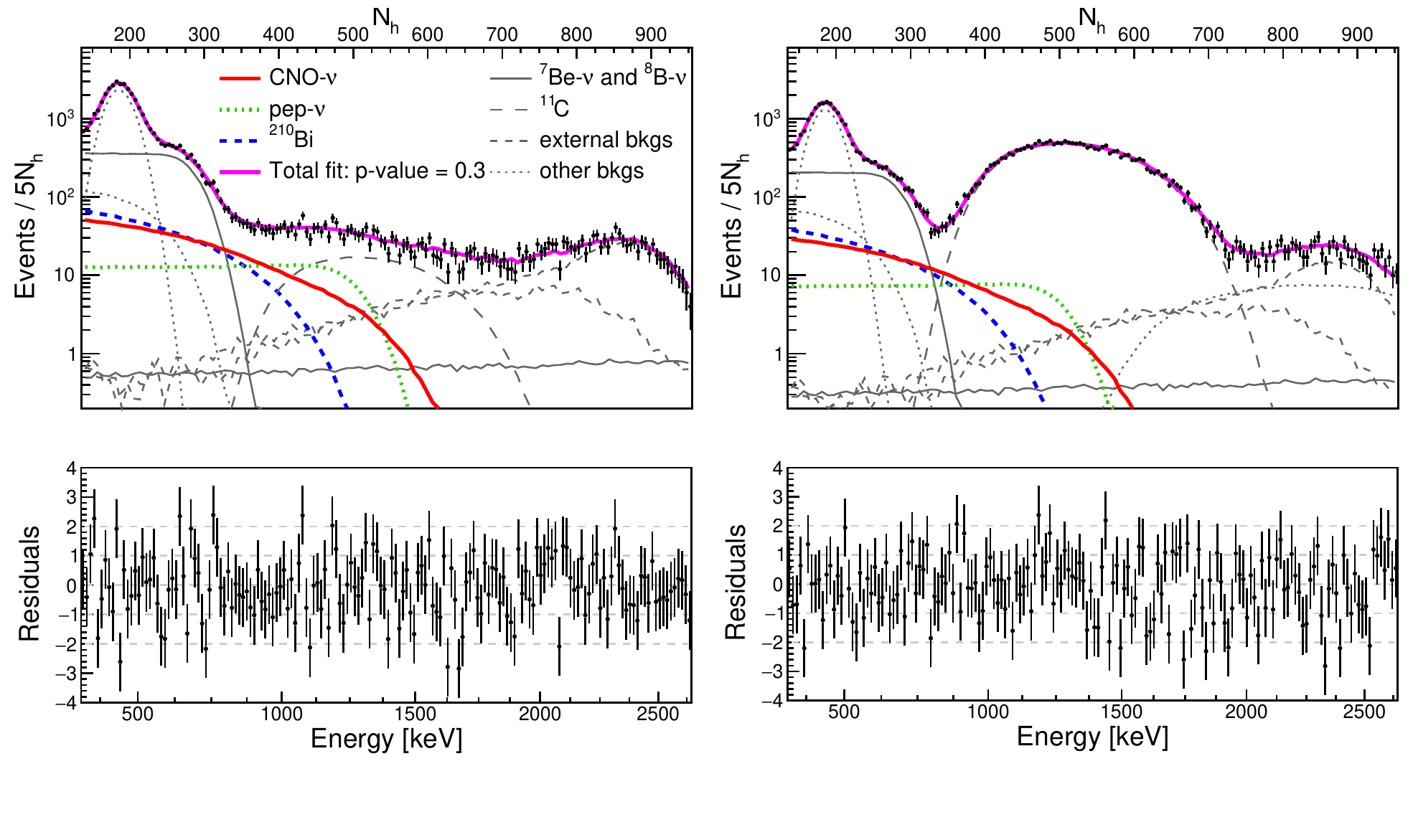}
\caption{\textsc{ED-Figure-8}}
\end{figure}

\begin{figure}
\centering
\includegraphics[width=\textwidth]{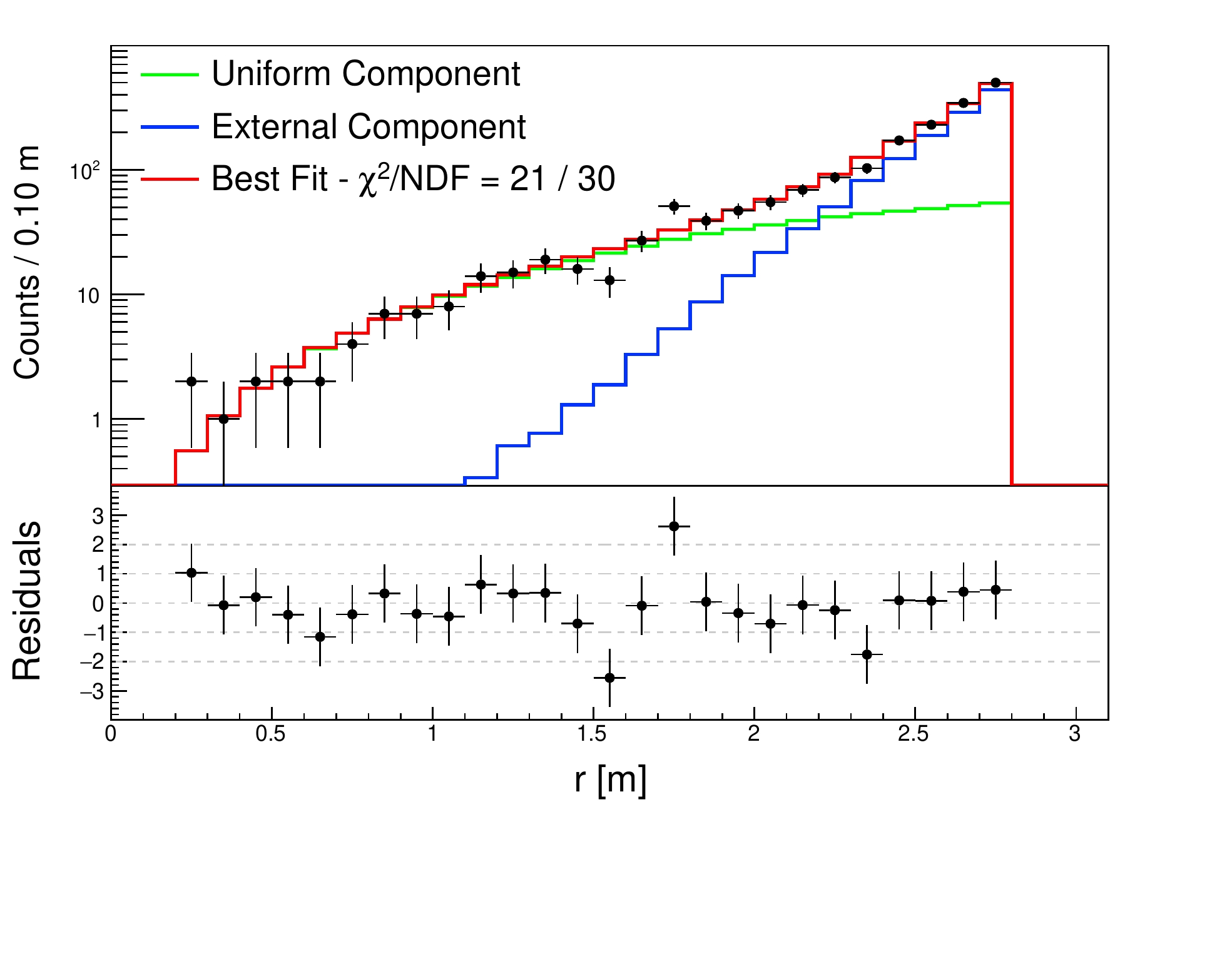}
\caption{\textsc{ED-Figure-9}}
\end{figure}

\begin{figure}
\centering
\includegraphics[width=\textwidth]{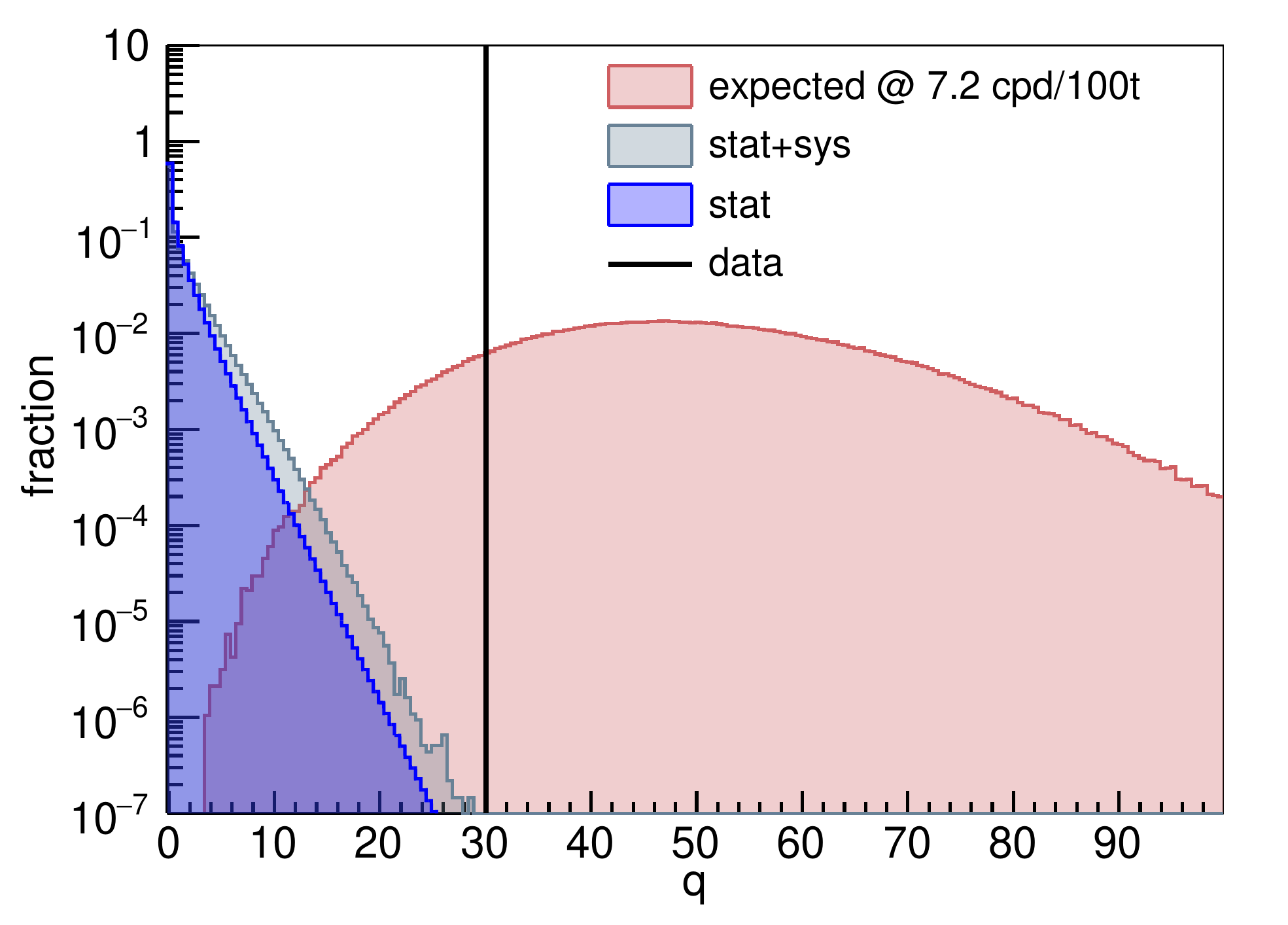}
\caption{\textsc{ED-Figure-10}}
\end{figure}


\begin{thebibliography}{1}

\bibitem{bib:SSM1}
J.N. Bahcall,
\newblock {Neutrino Astrophysics},
\newblock {\em Cambridge University Press}, 1989.

\bibitem{Vinyoles:2016djt}
N.~Vinyoles, A.M. Serenelli, F.L. Villante, S.~Basu, J.~Bergström, M.C.
  Gonzalez-Garcia, M.~Maltoni, C.~Peña-Garay, and N.~Song, A New Generation of Standard Solar Models,
\newblock {\em Astrophys. J.}, 835(2):202, 2017.

\bibitem{salaris}
M.~Salaris and S.~Cassisi,
\newblock {Evolution of Stars and Stellar Populations},
\newblock {\em John Wiley \& Sons Ltd.}, 2005.

\bibitem{Angulo1999}
C.~Angulo et~al., A compilation of charged-particle induced thermonuclear reaction rates, 
\newblock {\em Nuclear Physics A}, 656(1):3--183, 1999.

\bibitem{bib:Homestake}  R. Davis, A Half-Century with Solar Neutrinos,  {\em Nobel Prize Lecture, www.nobelprize.org}, 2002.

\bibitem{bib:Gallex} P. Anselmann et al. (GALLEX collaboration), Solar neutrinos observed by GALLEX at Gran Sasso {\em Phys. Lett. B 285, 376}, 1992.

\bibitem{bib:SAGE} J. Abdurashitov et al. (SAGE collaboration), Results from SAGE (The Russian-American Gallium solar neutrino experiment), {\em Phys. Lett. B 328, 234}, 1994.

\bibitem{bib:Art-Nobel2015}
A.B. McDonald, The Sudbury Neutrino Observatory: Observation of Flavor Change for Solar Neutrinos, {\em Nobel Prize Lecture, www.nobelprize.org}, 2015.

\bibitem{bib:SK} K. Hirata et al. (Kamiokande-II collaboration), Observation of $^8B$ solar neutrinos in the Kamiokande-II detector, {\em Phys. Rev. Lett. 63, 16}, 1989.

\bibitem{bib:SNO} Q. Ahmad et al. (SNO collaboration), Direct Evidence for Neutrino Flavor Transformation from Neutral-Current Interactions in the Sudbury Neutrino Observatory, {\em Phys. Rev. Lett. 89, 011301}, 2002.

\bibitem{bib:KamLAND}  T. Arakiet al. (KamLAND collaboration), Measurement of Neutrino Oscillation with KamLAND: Evidence of Spectral Distortion, {\em Phys. Rev. Lett. 94, 081801}, 2005.

\bibitem{bib:Nature-2018}
M.Agostini et~al. (Borexino Collaboration), Comprehensive measurement of $pp$-chain solar neutrinos,
\newblock {\em Nature}, 562(7728):505--510, 2018.

\bibitem{bib:Bethe1939}
H.A. Bethe., Energy production in stars, 
\newblock {\em Physical Review}, 55(5):434--456, 1939.

\bibitem{bib:Weizsacker}
C.F. Weizs\"acker, On Elementary Transmutations in the Interior of Stars: Paper II,
\newblock {\em Physik. Zeit.}, (38), 1937.

\bibitem{Bahcall1990}
J.N. Bahcall, Line versus continuum solar neutrinos, 
\newblock {\em Physical Review D}, 41(10):2964--2966, 1990.

\bibitem{Stonehill2004}
L.C. Stonehill, J.~A. Formaggio, and R.~G.~H. Robertson, Solar neutrinos from CNO electron capture,
\newblock {\em Physical Review C}, 69(1), 2004.

\bibitem{Villante2015}
F.L. Villante, CNO solar neutrinos: A challenge for gigantic ultra-pure liquid scintillator detectors, 
\newblock {\em Physics Letters B}, 742:279--284, 2015.

\bibitem{bib:LowZHighZ}
A.M. Serenelli, W.~C. Haxton, and C.~Pe{\~{n}}a-Garay, Solar Models with Accretion. I. Application to the Solar Abundance Problem,
\newblock {\em The Astrophysical Journal}, 743(1):24, 2011.

\bibitem{bib:DetPaper}
G.Alimonti et~al. (Borexino Collaboration), The Borexino detector at the Laboratori Nazionali del Gran Sasso, 
\newblock {\em Nuclear Instruments and Methods in Physics Research Section A:
  Accelerators, Spectrometers, Detectors and Associated Equipment},
  600(3):568--593, 2009.

\bibitem{bib:be7Long}
G.~Bellini et~al. (Borexino Collaboration), Final results of Borexino Phase-I on low-energy solar neutrino spectroscopy, 
\newblock {\em Phys. Rev. D}, 89(11):112007, 2014.

\bibitem{bib:Nusol}
M.~Agostini et~al. (Borexino Collaboration), Simultaneous precision spectroscopy of $pp$, $^7$Be and $pep$ 
 solar neutrinos with Borexino Phase-II,
\newblock {\em Physical Review D}, 100(8), 2019.

\bibitem{bib:ScienceTechnology2002}
G.~Alimonti et~al. (Borexino Collaboration), Science and Technology of BOREXINO: A Real Time Detector for Low Energy Solar Neutrinos,
\newblock {\em Astrop. Phys.}, 16:205--2034, 2002.

\bibitem{bib:Nature-2014}
M.~Agostini et~al. (Borexino Collaboration), Neutrinos from the primary proton–proton fusion process in the Sun, 
\newblock {\em Nature}, 512(7515):383--386, 2014.

\bibitem{bib:Be7-2011}
G.~Bellini et~al. (Borexino Collaboration), Precision Measurement of the $^7$Be Solar Neutrino Interaction Rate in Borexino,
\newblock {\em Physical Review Letters}, 107(14), 2011.

\bibitem{geonu2019}
M.~Agostini et~al. (Borexino Collaboration), Comprehensive geoneutrino analysis with Borexino,
\newblock {\em Phys. Rev. D}, 101:012009, 2020.

\bibitem{bib:xfd}
X.F. Ding, GooStats: A GPU-based framework for multi-variate analysis in particle physics, 
\newblock {\em Journal of Instrumentation}, 13(12):P12018--P12018, 2018.

\bibitem{bib:sensitivity-paper}
M.~Agostini et~al. (Borexino Collaboration), Sensitivity to neutrinos from the solar CNO cycle in Borexino, 
\newblock arXiv:2005.12829, 2020.

\bibitem{bib:Vissani2019}
F.~Vissani,
\newblock Solar Neutrinos, {\em World Scientific} , pp. 121-141 (2019).

\bibitem{bib:bergstrom}
J.~Bergstr{\"o}m, M.C. Gonzalez-Garcia, M.~Maltoni, C.~Pe{\~{n}}a-Garay, A.M.
  Serenelli, and N.~Song, Updated determination of the solar neutrino fluxes from solar neutrino data,
\newblock {\em JHEP 03}, 2016:132, 2016.

\bibitem{bib:NuPars}
F.~Capozzi, E.~Lisi, A.~Marrone, and A.~Palazzo, Global analysis of oscillation parameters, 
\newblock {\em J.Phys.Conf.Ser.}, 1312, 2019.

\bibitem{Villante2011}
F.L. Villante, A.~Ianni, F.~Lombardi, G.~Pagliaroli, and F.~Vissani, A step toward CNO solar neutrino detection in liquid scintillators, 
\newblock {\em Physics Letters B}, 701(3):336--341, 2011.

\bibitem{bib:diff}
M.~Wojcik, W.~Wlazlo, G.~Zuzel, and G.~Heusser, Radon diffusion through polymer membranes used in the solar neutrino experiment Borexino,
\newblock {\em Nucl. Instrum. Meth. A}, 449:158--171, 2000.

\bibitem{BravoBerguo2018}
D.~Bravo-Bergu{\~{n}}o et~al., 
The Borexino Thermal Monitoring \& Management System and simulations of the fluid-dynamics of the Borexino detector under asymmetrical, changing boundary conditions,
\newblock {\em Nucl. Instrum. Meth. A}, 885:38--53, 2018.

\bibitem{bib:PoSimulation}
V.~Di~Marcello, D.~Bravo-Berguño, R.~Mereu, F.~Calaprice, A.~Di~Giacinto,
  A.~Di~Ludovico, Aldo Ianni, Andrea Ianni, N.~Rossi, and L.~Pietrofaccia, Fluid-dynamics and transport of 210Po in the scintillator Borexino detector: A numerical analysis, 
\newblock {\em Nucl. Instrum. Meth. A}, 964:163801, 2020.

\bibitem{bib:healpix}
K.M. Gorski, B.D. Wandelt, F.K. Hansen, E. Hivon, and A.J. Banday. The HEALPix Primer.
\newblock arXiv:9905275, 1999.


\bibitem{bib:MCPaper}
M.~Agostini et~al. (Borexino Collaboration), The Monte Carlo simulation of the Borexino detector,
\newblock {\em Astropart. Phys.}, 97:136--159, 2018.

\bibitem{bib:Daniel1962}
H.~Daniel, Das $\beta$-spektrum des RaE,
\newblock {\em Nuclear Physics}, 31:293--307, 1962.

\bibitem{bib:Carles1996}
A. Grau Carles and A. Grau Malonda, Precision measurement of the RaE shape factor, 
\newblock {\em Nuclear Physics A}, 596(1):83--90, 1996.

\bibitem{bib:210BiDerbin} 
I.E.~Alekseev et~al, Precision measurement \bi $\beta$-spectrum,
\newblock e-print: 2005.08481.

\bibitem{bib:bxcalib}
H.~Back et~al. (Borexino Collaboration), Borexino calibrations: Hardware, Methods, and Results,
\newblock {\em JINST}, 7:P10018, 2012.

\bibitem{deHolanda2004}
P.C. de~Holanda, W.~Liao, and A.Yu. Smirnov, Toward precision measurements in solar neutrinos,
\newblock {\em Nuclear Physics B}, 702(1-2):307--332, 2004.

\bibitem{Capozzi2018}
F.~Capozzi, E.~Lisi, A.~Marrone, and A.~Palazzo, Current unknowns in the three neutrino framework,
\newblock {\em Progress in Particle and Nuclear Physics}, 102:48--72, 2018.

\bibitem{bib:cowan}
G.~Cowan, K.~Cranmer, E.~Gross, and O.~Vitells, 
Asymptotic formulae for likelihood-based tests of new physics,
\newblock {\em Eur. Phys. J. C}, 71:1554, 2011.
\newblock (Erratum: {\em Eur. Phys. J. C} 73, 2501, 2013).

\bibitem{jureca}
 D. Krause and P. Th\"ornig, JURECA, Modular supercomputer at jülich supercomputing centre,
\newblock {\em Journal of large-scale research facilities}, 4(A132), 2018.

\bibitem{birks64}
J.B. Birks, The Theory and practice of scintillation counting,
\newblock {\em Pergamon Press, Oxford}, 1964.

\bibitem{bib:PurifPlants}
J.~Benziger et~al, The Scintillator Purification System for the Borexino Solar Neutrino Detector
\newblock {\em Nucl. Instrum. Meth. A}, 587:277--291, 2008.

\bibitem{bib:FluidHandling}
G.~Alimonti et~al. (Borexino Collaboration), The liquid handling systems for the Borexino solar neutrino detector,
\newblock {\em Nucl. Instrum. Meth. A}, 609:58--78, 2009.

\bibitem{bib:Muons}
G.~Bellini et~al. (Borexino Collaboration), Cosmic-muon flux and annual modulation in Borexino at 3800 m water-equivalent depth,
\newblock {\em JCAP}, 05:015, 2012.

\bibitem{bib:Cosmogenics}
G.~Bellini et~al. (Borexino Collaboration), Cosmogenic Backgrounds in Borexino at 3800 m water-equivalent depth,
\newblock {\em JCAP}, 08:049, 2013.

\bibitem{bib:Neutrons}
G.~Bellini et~al. (Borexino Collaboration), Muon and cosmogenic neutron detection in Borexino,
\newblock {\em JINST}, 6:P05005, 2011.

\bibitem{bib:stokes}
C.~Miller~Cruickshank, 
The Stokes-Einstein law for diffusion in solution,
\newblock {\em Royal Society}, 106, 1924.

\bibitem{bib:tmva}
A.~Hoecker, P.~Speckmayer, J.~Stelzer, J.~Therhaag, H.~von Toerne and E. ~Voss,
TMVA - Toolkit for Multivariate Data Analysis,
\newblock {\em PoS}, ACAT:040, 2007.

\bibitem{bib:multinest1}
F.~Feroz, M.P. Hobson, E.~Cameron, and A.N. Pettitt, Importance Nested Sampling and the MultiNest Algorithm, 
\newblock {\em Open J.Astrophys. 2-1, 10}, 2019. 

\bibitem{bib:multinest2}
F.~Feroz, M.~P. Hobson, and M.~Bridges, MultiNest: an efficient and robust Bayesian inference tool for cosmology and particle physics,
\newblock {\em Monthly Notices of the Royal Astronomical Society},
  398(4):1601--1614, 2009.

\bibitem{bib:multinest3}
F.~Feroz and M.~P. Hobson, Multimodal nested sampling: an efficient and robust alternative to Markov Chain Monte Carlo methods for astronomical data analyses,
\newblock {\em Monthly Notices of the Royal Astronomical Society},
  384(2):449--463, 2008.

\bibitem{bib:fick}
A. Fick, Ueber Diffusion,
\newblock {\em Annalen der Physik}, 170(1):59--86, 1855.

\bibitem{bib:seas}
M.~Agostini et~al. (Borexino Collaboration), 
Seasonal modulation of the $^7$Be solar neutrino rate in Borexino,
\newblock {\em Astropart. Phys.}, 92:21--29, 2017.

\bibitem{bib:pep}
G.~Bellini et~al. (Borexino Collaboration), First Evidence of pep Solar Neutrinos by Direct Detection in Borexino,
\newblock {\em Phys. Rev. Lett.}, 108:051302, 2012.

\bibitem{bib:cousin}
R.D. Cousins and V.L. Highland, Incorporating systematic uncertainties into an upper limit,
\newblock {\em Nucl. Instrum. Meth. A}, 320:331--335, 1992.

\bibitem{bib:jeffery}
L.D. Brown, T.T. Cai, and A.~Das Gupta, Interval Estimation for a Binomial Proportion,
\newblock {\em Statistical science}, 16(2):101--133, 2001.

\end{thebibliography}
\end{document}